\documentclass[journal]{IEEEtran}
\usepackage{graphicx}
\graphicspath{{figures/}}
\usepackage{amsmath,amssymb,dsfont,verbatim}
\usepackage{array}
\usepackage[caption=false,font=footnotesize]{subfig}
\usepackage{import}
\usepackage{url, hyperref}
\usepackage{cite}
\usepackage{xcolor}
\usepackage{booktabs} \newcommand{\ra}[1]{\renewcommand{\arraystretch}{#1}}
\usepackage{tikz}
\usetikzlibrary{mindmap}

\newif\ifarx
 \arxtrue 

\usepackage{amsthm}

\DeclareMathOperator{\cw}{{\scriptstyle\mathcal{W}}}
\DeclareMathOperator{\bcw}{{\boldsymbol{\scriptstyle\mathcal{W}}}}
\DeclareMathOperator*{\argmin}{arg\,min}

\DeclareMathOperator{\T}{\mathsf{T}}
\DeclareMathOperator{\E}{\mathds{E}}

\DeclareMathOperator{\w}{\boldsymbol{w}}
\DeclareMathOperator{\x}{\boldsymbol{x}}
\DeclareMathOperator{\s}{\boldsymbol{s}}

\DeclareMathOperator{\h}{\boldsymbol{h}}

\usepackage{amsthm}

\newtheorem{definition}{Definition}
\newtheorem{assumption}{Assumption}
\newtheorem{theorem}{Theorem}

\newtheorem{lemma}{Lemma}

\begin{document}%
\title{Distributed Learning in Non-Convex Environments -- Part II: Polynomial Escape from Saddle-Points }%
\author{Stefan Vlaski,~\IEEEmembership{Student Member,~IEEE,}
 				and Ali H. Sayed,~\IEEEmembership{Fellow,~IEEE}
\thanks{The authors are with the Institute of Electrical Engineering, \'{E}cole Polytechnique F\'{e}d\'{e}rale de Lausanne. S. Vlaski is also with the Department of Electrical Engineering, University of California, Los Angeles. This work was supported in part by NSF grant CCF-1524250. Emails:\{stefan.vlaski, ali.sayed\}@epfl.ch.}\thanks{A limited short version of this work appears in the conference publication~\cite{Vlaski19}.}}%
\maketitle
\begin{abstract}
The diffusion strategy for distributed learning from streaming data employs local stochastic gradient updates along with exchange of iterates over neighborhoods. In Part I~\cite{Vlaski19nonconvexP1} of this work we established that agents cluster around a network centroid and proceeded to study the dynamics of this point. We established expected descent in non-convex environments in the large-gradient regime and introduced a short-term model to examine the dynamics over finite-time horizons. Using this model, we establish in this work that the diffusion strategy is able to escape from strict saddle-points in \( O(1/\mu) \) iterations; it is also able to return approximately second-order stationary points in a polynomial number of iterations. Relative to prior works on the {polynomial} escape from saddle-points, most of which focus on \emph{centralized} perturbed or stochastic gradient descent, our approach requires less restrictive conditions on the gradient noise process.
\end{abstract}
\section{Introduction}\label{sec:intro}
\IEEEPARstart{W}{e} consider a network of \( N \) agents. Each agent \( k \) is equipped with a local, stochastic cost of the form \( J_k(w) = \E_x Q_k(w; \x_k) \), where \( w \in \mathds{R}^M \) denotes a parameter vector and \( \x_k \) denotes random data. In Part I~\cite{Vlaski19nonconvexP1}, we consider a global optimization problem of the form:
\begin{equation}\label{eq:global_problem}
	\min_w J(w),\ \ \ \ \ \ \ \ \mathrm{where}\ J(w) \triangleq \sum_{k=1}^{N} p_k J_k(w)
\end{equation}
where the weights \( p_k \) are a function of the combination weights \( a_{\ell k} \) and will be specified further below in~\eqref{eq:perron}.

Solutions to such problems via distributed strategies can be pursued through a variety of algorithms, including those of the consensus and diffusion type~\cite{Nedic09, Sayed14, Chen15transient, Xin19, Yuan16, Shi15, Yuan18}. In Part I~\cite{Vlaski19nonconvexP1}, we studied the diffusion strategy strategy due to its proven enhanced performance in adaptive environments in response to streaming data and drifting conditions~\cite{Tu12, Sayed14}. The strategy takes the form:
\begin{subequations}
\begin{align}
  \boldsymbol{\phi}_{k,i} &= \w_{k,i-1} - \mu \widehat{\nabla J}_{k}(\w_{k,i-1})\label{eq:adapt}\\
  \w_{k,i} &= \sum_{\ell=1}^{N} a_{\ell k} \boldsymbol{\phi}_{\ell,i}\label{eq:combine}
\end{align}
\end{subequations}
Note that the gradient step~\eqref{eq:adapt} employs a \emph{stochastic} gradient approximation \( \widehat{\nabla J}_{k}(\w_{k,i-1}) \), rather than the true gradient \( {\nabla J}_{k}(\w_{k,i-1}) \). The random approximation of the true gradient based on sampled data introduces persistent gradient noise, which seeps into the evolution of the algorithm. A commonly employed construction is \( \widehat{\nabla J}_{k}(\w_{k,i-1}) = {\nabla Q}_{k}(\w_{k,i-1}; \x_k) \){; nevertheless, we consider general stochastic gradient approximations \( \widehat{\nabla J}_{k}(\w_{k,i-1}) \) under suitable conditions on the induced gradient noise process (Assumptions~\ref{as:gradientnoise} and~\ref{as:noise_in_saddle} further ahead)}. Prior works have studied the dynamics of the diffusion strategy~\eqref{eq:adapt}--\eqref{eq:combine} and examined the implications of the gradient noise term in the \emph{{strongly}-convex} setting~\cite{Sayed14, Chen15transient, Chen15performance}. In particular, it has been shown that despite the presence of gradient noise, the iterates \( \w_{k, i} \) will approach the global solution \( w^{\star} \triangleq \argmin_w J(w) \) to the problem~\eqref{eq:global_problem} in the mean-square-error sense, namely it will hold that \( \limsup_{i \to \infty} \E {\| w^{\star} - \w_{k, i} \|}^2 = O(\mu) \).

In Part I~\cite{Vlaski19nonconvexP1} we showed that many of the desirable properties of the diffusion algorithm continue to hold in the more challenging non-convex setting. We established that all agents will cluster around a common network centroid after sufficient iterations and established expected descent of the network centroid in the large-gradient regime. In this part of the work we establish that the diffusion strategy is able to escape strict-saddle points and return second-order stationary points in polynomial time.

\subsection{Related Works}
A general discussion on decentralized algorithms for optimization and learning~\cite{Nedic09, Sayed14, Chen15transient, Xin19, Yuan16, Shi15, Yuan18, Nassif16, Towfic15, Ying18} can be found in Part I~\cite{Vlaski19nonconvexP1}. In this section, we focus on works studying the ability of algorithms to escape strict saddle-points and reach second-order stationary points, which is the focus of this part. The desire to obtain guarantees for the escape from saddle-points is motivated by the observation that in many problems of interest, such as neural networks, saddle-points can correspond to bottlenecks of the optimization problem. As such, guarantees of convergence to first-order stationary points, i.e., points where the norm of the gradient is small, need not be sufficient to establish good performance. For this reason, there has been interest in the guarantee of convergence to second-order stationary points. Approximate second-order stationary points, like first-order stationary points, are required to have a small gradient norm, but are also restricted in terms of the smallest eigenvalues of their Hessian matrices.

Works that study the ability of gradient descent algorithms to escape strict saddle-points {can broadly be classified into two approaches. The first class is based on the fact that there is at least one direction of descent at every saddle-point and leverage either second-order information~\cite{Nesterov06} or first-order strategies for identifying a negative-curvature direction~\cite{Fang18, Allen18neon, Allen18natasha} to identify the descent direction. Our work falls into a second class of strategies, which exploit the fact that \emph{strict} saddle-points (defined later) are unstable in the sense that small perturbations allow for the iterates to escape from the saddle point almost surely.} Along these lines, it has been shown in~\cite{Lee16} that under an appropriately chose random initialization scheme, the gradient descent algorithm converges to minimizers almost surely. The work~\cite{Scutari18} further leveraged this fact to establish that distributed gradient descent with appropriately chosen initialization escapes saddle points. When subjected to persistent, but diminishing perturbations, known as annealing, \emph{asymptotic} almost sure convergence to global minimizers of gradient descent-type algorithms has also been established in the centralized~\cite{Gelfand91} and more recently in the distributed setting~\cite{Swenson19}. All these useful results, while powerful in theory, still do not provide a guarantee that the procedures are efficient in the sense that they would return accurate solutions after a finite number of iterations. Actually, despite the fact that gradient descent with random initialization escapes saddle-points almost surely~\cite{Lee16}, it has been established that this process can take exponentially long~\cite{Du17}, rendering the procedure impractical.

These observations have sparked interest in the design of methods that have the ability to escape saddle-points \emph{efficiently}, where efficiency is loosely defined as yielding success in polynomial, rather than exponential time. The authors in~\cite{Ge15} add persistent, i.i.d. perturbations to the exact gradient descent algorithm and establish polynomial escape from saddle-points, while the work~\cite{Jin17} adds perturbations only when the presence of a saddle-point is detected. It is important to note that in most of these works, perturbations or random initializations are selected and introduced with the explicit purpose of allowing the algorithm to escape from unstable stationary points. For example, random initialization is followed by exact gradient updates in the works~\cite{Lee16, Scutari18}, while the perturbations in~\cite{Jin17} are applied only when a saddle-point is detected via the norm of the gradient. All of these techniques still require knowledge of the \emph{exact gradient}. While the authors of~\cite{Ge15} consider persistent gradient perturbations, these are nevertheless assumed to be independentently and identically distributed.

Motivated by these considerations, in this work, we focus on implementations that employ \emph{stochastic} gradient approximations and \emph{constant} step-sizes. This is driven by the fact that computation of the exact gradients \( \nabla J_k(\cdot) \) is generally infeasible in practice because (a) data may be streaming in, making it impossible to compute \( \nabla \E_{x_k} Q_k(\cdot; \x_k) \) in the absence of knowledge about the distribution of the data or (b) the data set, while available as a batch, may be so large that efficient computation of the full gradient is infeasible. As such, the exact gradient will need to be replaced by an approximate \emph{stochastic} gradient, which ends up introducing in a natural manner some form of \emph{gradient noise} into the operation of the algorithm; this noise is the difference between the true gradient and its approximation. The gradient noise seeps into the operation of the algorithm continually and becomes coupled with the evolution of the iterates, resulting in perturbations that are neither identically nor independently distributed over time. For instance, the presence of the gradient noise process complicates the dynamics of the iterate evolution relative to the centralized recursions considered in~\cite{Ge15}.

There have been some recent works that study {\em stochastic} gradient scenarios as well. However, these methods alter the gradient updates in specific ways or require the gradient noise to satisfy particular conditions. For example, the work~\cite{Jin19} proposes the addition of Gaussian noise to the naturally occuring gradient noise, while the authors of~\cite{HadiDaneshmand18} leverage alternating step-sizes. The works~\cite{Fang18, Allen18neon, Allen18natasha} introduce an intermediate negative-curvature-search step. All of these works alter the traditional stochastic gradient algorithm in order to ensure efficient escape from saddle-points. The work~\cite{Fang19} studies the traditional stochastic gradient algorithm under a dispersive noise assumption.

{The key contributions of this work are three-fold. To the best of our knowledge, we present the first analysis establishing \emph{efficient} (i.e., polynomial) escape from strict-staddle points in the \emph{distributed} setting. Second, we establish that the gradient noise process is sufficient to ensure efficient escape without the need to alter it by adding artificial forms of perturbations{, interlacing steps with small and large step-sizes or imposing a dispersive noise assumption, as long as there is a gradient noise component present in some descent direction for every strict saddle-point.} Third, relative to the existing literature on \emph{centralized} non-convex optimization, where the focus is mostly on deterministic or \emph{finite-sum} optimization, our modeling conditions are specifically tailored to the scenario of learning from stochastic \emph{streaming} data. In particular, we only impose bounds on the gradient noise variance in expectation, rather than assume a bound with probability \( 1 \)~\cite{HadiDaneshmand18, Fang19} or a sub-Gaussian distribution~\cite{Jin19}. Furthermore, we assume that any Lipschitz conditions only hold on the \emph{expected} stochastic gradient approximation, rather than for every realization, with probability \( 1 \)~\cite{Fang18, Allen18neon, Allen18natasha}.}

For ease of reference, the modeling conditions and results from this and related works are summarized in Table~\ref{tab:references}.

\begin{table*}\centering
\ra{1.3}
\begin{tabular}{@{}ccccccccc@{}}\toprule
& \multicolumn{5}{c}{Modeling conditions} & \phantom{abc}& \multicolumn{2}{c}{Results}\\ \cmidrule{2-6} \cmidrule{8-9}
& Gradient & Hessian & Initialization & Perturbations & Step-size && Stationary & Saddle \\ \midrule
\textbf{Centralized}\\
~\cite{Gelfand91}  & Lipschitz & --- & --- & SGD + Annealing & diminishing && \( \checkmark \) & asymptotic\(^{\dagger}\) \\
~\cite{Ge15}  & Lipschitz \& bounded\(^{\star}\) & Lipschitz & --- & i.i.d.\ and bounded w.p. 1 & constant && \( \checkmark \) & polynomial \\
~\cite{Lee16}  & Lipschitz & --- & Random & --- & constant && \( \checkmark \) & asymptotic \\
~\cite{Jin17}  & Lipschitz & Lipschitz & --- & Selective \& bounded w.p. 1 & constant && \( \checkmark \) & polynomial \\
~\cite{HadiDaneshmand18}  & Lipschitz & Lipschitz & --- & SGD, bounded w.p. 1 & alternating && \( \checkmark \) & polynomial \\
~\cite{Fang18}  & Lipschitz & Lipschitz & --- & Bounded variance, Lipschitz w.p. 1 & constant && \( \checkmark \) & polynomial \\
~\cite{Allen18natasha}  & Lipschitz & Lipschitz & --- & Bounded variance, Lipschitz w.p. 1 & constant && \( \checkmark \) & polynomial \\
~\cite{Allen18neon}  & Lipschitz & Lipschitz & --- & Bounded variance, Lipschitz w.p. 1 & constant && \( \checkmark \) & polynomial \\
~\cite{Fang19}  & Lipschitz & Lipschitz & --- & SGD, bounded w.p. 1 & constant && \( \checkmark \) & polynomial \\
~\cite{Jin19}  & Lipschitz & Lipschitz & --- & SGD + Gaussian &constant && \( \checkmark \)  & polynomial \\
\textbf{Decentralized}\\
~\cite{Lorenzo16} & Lipschitz \& bounded & --- & --- & ---&constant && \( \checkmark \) & ---\\
~\cite{Wang18} & Lipschitz & --- & --- & ---&constant && \( \checkmark \) & ---\\
~\cite{Tatarenko17} & Lipschitz \& bounded & --- & --- & i.i.d.&diminishing && \( \checkmark \) & ---\\
~\cite{Scutari18} & Lipschitz & Exists & Random & --- &constant && \( \checkmark \) & asymptotic\\
~\cite{Swenson19}  & Bounded disagreement & --- & --- & SGD + Annealing & diminishing && \( \checkmark \) & asymptotic\(^{\dagger}\) \\
\textbf{This work} & \textbf{Bounded disagreement} & \textbf{Lipschitz} & \textbf{---} & \textbf{Bounded moments} &\textbf{constant} && \( \boldsymbol{\checkmark} \) & \textbf{polynomial} \\ \bottomrule\
\end{tabular}
\caption{Comparison of modeling assumptions and results for gradient-based methods. Statements marked with \(^{\star}\) are not explicitly stated but are implied by other conditions. The works marked with \(^{\dagger}\) establish global (asymptotic) convergence, which of course implies escape from saddle-points.}\label{tab:references}
\end{table*}

\section{Review of Part I~\cite{Vlaski19nonconvexP1}}
\subsection{Modeling Conditions}
In this section, we briefly list the modeling conditions employed in Part I~\cite{Vlaski19nonconvexP1}. For a more detailed discussion, we refer the reader to~\cite{Vlaski19nonconvexP1}.
\begin{assumption}[\textbf{Strongly-connected graph}]\label{as:strongly_connected}
  The combination weights in~\eqref{eq:combine} are convex combination weights satisfying:
  \begin{equation}\label{eq:combinationcoef}
    a_{\ell k} \geq 0, \quad \sum_{\ell \in \mathcal{N}_k} a_{\ell k}=1, \quad a_{\ell k} = 0\ \mathrm{if}\ \ell \notin \mathcal{N}_k
  \end{equation}
  The symbol \( {\cal N}_k \) denotes the set of neighbors of agent \( k \). We shall assume that the graph described by the weighted combination matrix \(A=[a_{\ell k}]\) is strongly-connected~\cite{Sayed14}. This means that there exists a path with nonzero weights between any two agents in the network and, moreover, at least one agent has a nontrivial self-loop, \(a_{kk}>0\).\hfill\IEEEQED%
\end{assumption}
The Perron-Frobenius theorem~\cite{Horn03,Pillai05,Sayed14} then implies that \( A \) has a spectral radius of one and a single eigenvalue at one. The corresponding eigenvector can be normalized to satisfy:
\begin{equation}\label{eq:perron}
  Ap=p, \quad \mathds{1}^{\T} p=1, \quad p_k > 0
\end{equation}
where the \( \{ p_k \} \) denote the individual entries of the Perron vector, \(p\).
\begin{assumption}[\textbf{Lipschitz gradients}]\label{as:lipschitz}
  For each \( k \), the gradient \( \nabla J_k(\cdot) \) is Lipschitz, namely, for any \( x,y \in \mathds{R}^{M} \):
  \begin{equation}\label{eq:lipschitz}
    \|\nabla J_k(x) - \nabla J_k(y)\| \le \delta \|x-y\|
  \end{equation}
	In light of~\eqref{eq:global_problem} and Jensen's inequality, this implies for the aggregate cost:
	\begin{equation}\label{eq:lipschitz_global}
		\|\nabla J(x) - \nabla J(y)\| \le \delta \|x-y\|
	\end{equation}
\end{assumption}\hfill\IEEEQED%

{\noindent The Lipschitz gradient conditions~\eqref{eq:lipschitz} and~\eqref{eq:lipschitz_global} imply
\begin{align}
  J(y) \le J(x) + {\nabla J(x)}^{\T} \left( y-x \right)  + \frac{\delta}{2} {\|x-y\|}^2 \label{eq:quadratic_upper}
\end{align}
For the Hessian matrix we have~\cite{Sayed14}:
\begin{align}\label{eq:hessian_bound}
  - \delta I \le \nabla^2 J(x) \le \delta I
\end{align}
}
{
\begin{assumption}[\textbf{Bounded gradient disagreement}]\label{as:bounded}
  For each pair of agents \( k \) and \( \ell \), the gradient disagreement is bounded, namely, for any \( x \in \mathds{R}^{M} \):
  \begin{equation}\label{eq:bounded}
    \|\nabla J_k(x) - \nabla J_{\ell}(x)\| \le G
  \end{equation}
\end{assumption}\hfill\IEEEQED}
\begin{definition}[\textbf{Filtration}]\label{def:filtration}
  We denote by \( \boldsymbol{\mathcal{F}}_{i} \) the filtration generated by the random processes \( \w_{k, j} \) for all \( k \) and \( j \le i \):
  \begin{equation}
    \boldsymbol{\mathcal{F}}_{i} \triangleq \left \{ \bcw_{0}, \bcw_{1}, \ldots, \bcw_{i} \right \}
  \end{equation}
  where \( \bcw_{j} \triangleq \mathrm{col}\left \{ \w_{1, j}, \ldots, \w_{k, j} \right \} \) contains the iterates across the network at time \( j \). Informally, \( \boldsymbol{\mathcal{F}}_{i} \) captures all information that is available about the stochastic processes \( \w_{k, j} \) across the network up to time \( i \).\hfill\IEEEQED
\end{definition}
\begin{assumption}[\textbf{Gradient noise process}]\label{as:gradientnoise}
  For each \( k \), the gradient noise process is defined as
  \begin{equation}
    \s_{k,i}(\w_{k,i-1}) = \widehat{\nabla J}_k(\w_{k,i-1}) - \nabla J_k(\w_{k,i-1})
  \end{equation}
  and satisfies
  \begin{subequations}
    \begin{align}
      \E \left\{ \s_{k,i}(\w_{k,i-1}) | \boldsymbol{\mathcal{F}}_{i-1} \right\} &= 0 \label{eq:conditional_zero_mean}\\
      \E \left\{ \|\s_{k,i}(\w_{k,i-1})\|^4 | \boldsymbol{\mathcal{F}}_{i-1} \right\} &\le \sigma^4 \label{eq:gradientnoise_fourth}
    \end{align}
  \end{subequations}
  for some non-negative constant \( \sigma^4 \). We also assume that the gradient noise pocesses are pairwise uncorrelated over the space conditioned on \( \boldsymbol{\mathcal{F}}_{i-1} \), i.e.:
  \begin{equation}
    \E \left\{ \s_{k,i}(\w_{k,i-1}) \s_{\ell,i}(\w_{\ell,i-1})^{\T} |  \boldsymbol{\mathcal{F}}_{i-1} \right\} = 0\label{eq:uncorrelated_noise}
  \end{equation}
  \hfill\IEEEQED%
\end{assumption}
The fourth-order condition also implies via Jensen's inequality:
\begin{align}
  &\: \E \left\{ \|\s_{k,i}(\w_{k,i-1})\|^2 | \boldsymbol{\mathcal{F}}_{i-1} \right\} \le \sigma^2 \label{eq:gradientnoise}
\end{align}
\begin{definition}[Sets]\label{DEF:SETS}
  To simplify the notation in the sequel, we introduce following sets:
  \begin{align}
    \mathcal{G} &\triangleq \left \{ w : {\left \| \nabla J(w) \right \|}^2 \ge \mu \frac{c_2}{c_1}\left(1+ \frac{1}{\pi}\right) \right \} \label{eq:define_g}\\
    \mathcal{G}^C &\triangleq \left \{ w : {\left \| \nabla J(w) \right \|}^2 < \mu \frac{c_2}{c_1} \left(1+\frac{1}{\pi} \right)\right \} \\
    \mathcal{H} &\triangleq \left \{ w : w \in \mathcal{G}^C, \lambda_{\min}\left( \nabla^2 J(w) \right) \le -\tau \right \} \label{eq:define_h}\\
    \mathcal{M} &\triangleq \left \{ w : w \in \mathcal{G}^C, \lambda_{\min}\left( \nabla^2 J(w) \right) > -\tau \right \} \label{eq:define_m}
  \end{align}
  where \( \tau \) is a small positive parameterm, \( c_1 \) and \( c_2 \) are constants:
  \begin{align}
		c_1 &\triangleq \frac{1}{2} \left(1 - 2 \mu \delta\right) = O(1) \label{eq:define_c1}\\
		c_2 &\triangleq \delta \sigma^2 / 2 = O(1) \label{eq:define_c2}
	\end{align}
  and \( 0 < \pi < 1 \) is a parameter to be chosen. Note that \( \mathcal{G}^C = \mathcal{H} \cup \mathcal{M} \). We also define the probabilities \(\pi^{\mathcal{G}}_i \triangleq \mathrm{Pr}\left \{ \w_{c, i} \in \mathcal{G} \right \}\), \( \pi^{\mathcal{H}}_i \triangleq \mathrm{Pr}\left \{ \w_{c, i} \in \mathcal{H} \right \}\) and \( \pi^{\mathcal{M}}_i \triangleq \mathrm{Pr}\left \{ \w_{c, i} \in \mathcal{M} \right \} \). Then for all \( i \), we have \( \pi^{\mathcal{G}}_i + \pi^{\mathcal{H}}_i + \pi^{\mathcal{M}}_i = 1 \). \hfill\IEEEQED
\end{definition}
\begin{assumption}[\textbf{Lipschitz Hessians}]\label{as:lipschitz_hessians}
  Each \( J_k(\cdot) \) is twice-differentiable with Hessian \( \nabla^2 J_k(\cdot) \) and, there exists \( \rho \ge 0 \) such that:
  \begin{equation}
    {\| \nabla^2 J_k(x) - \nabla^2 J_k(y) \|} \le \rho \|x - y\|
  \end{equation}
  By Jensen's inequality, this implies that \( J(\cdot) = \sum_{k=1}^N p_k J_k(\cdot) \) also satisfies:
  \begin{equation}\label{eq:lipschitz_hessians}
    {\| \nabla^2 J(x) - \nabla^2 J(y) \|} \le \rho \|x - y\|
  \end{equation}\hfill\IEEEQED
\end{assumption}
\noindent Similarly to the quadratic upper bound that follows from the Lipschitz condition on the first-derivative~\eqref{eq:quadratic_upper}, this new Lipschitz condition on the second-derivative implies a cubic upper bound on the function values~\cite{Nesterov06}:
\begin{align}
  J(y) \le&\: J(x) + {\nabla J(x)}^{\T} (y-x) + \frac{1}{2} {(y-x)}^{\T} \nabla^2 J(x) (y-x) \notag \\
  &\: + \frac{\rho}{6} {\left \| y-x \right\|}^3 \label{eq:cubic_upper}
\end{align}

\subsection{Review of Results}
An important quantity in the network dynamics of~\eqref{eq:adapt}--\eqref{eq:combine} is the weighted network centroid:
\begin{equation}
  \w_{c, i} \triangleq \sum_{k=1}^N p_k \w_{k, i}
\end{equation}
where the weights \( p_k \) are elements of the Perron vector, defined in~\eqref{eq:perron}, which in turn is a function of the graph topology and weights. The network centroid can be shown to evolve according to a perturbed, centralized, exact gradient descent recursion~\cite{Chen15transient}:
\begin{align}\label{eq:perturbed_gradient_descent}
	\w_{c, i} = \w_{c, i-1} - \mu \sum_{k=1}^N p_k {\nabla J}_k (\w_{c, i-1}) - \mu \boldsymbol{d}_{i-1} - \mu \s_i
\end{align}
where we defined the perturbation terms:
\begin{align}
	\boldsymbol{d}_{i-1} &\triangleq \sum_{k=1}^N p_k \left( {\nabla J}_k (\w_{k, i-1}) - {\nabla J}_k (\w_{c, i-1}) \right) \\
	\boldsymbol{s}_{i} &\triangleq \sum_{k=1}^N p_k \left( \widehat{\nabla J}_k (\w_{k, i-1}) - {\nabla J}_k (\w_{k, i-1}) \right) \label{eq:centralized_gradient_noise}
\end{align}
In Part I~\cite[Theorem 1]{Vlaski19nonconvexP1} we established that, under assumptions~\ref{as:strongly_connected}--\ref{as:gradientnoise}, all agents will cluster around the network centroid in the mean-fourth sense:
\begin{align}
	&\:\E {\left \| \bcw_i - \left( \mathds{1} p^{\T} \otimes I \right) \bcw_{i} \right \|}^4 \notag \\
  \le&\: \mu^4 {\left \| \mathcal{V}_L \right \|}^4 \frac{{\left \|J_{\epsilon}^{\T} \right \|}^4}{{\left(1-{\left \|J_{\epsilon}^{\T} \right \|}\right)}^4} {\| \mathcal{V}_R^{\T} \|}^4 N^2 \left( G^4 + \sigma^4 \right) + o(\mu^4)\label{eq:network_disagreement_fourth}
\end{align}
for \( i \ge i_o \) where \( i_o \triangleq {\log\left( o(\mu^4) \right)}/{\log\left( {\left \|J_{\epsilon}^{\T} \right \|} \right)} \). This result has two implications. First, it establishes that, despite the fact that agents may be descending along different cost functions, and despite the fact that they may have been initialized close to different local minima, the entire network will eventually agree on a common iterate in the mean-fourth sense (and via Markov's inequality with high probability). Furthermore, it allows us to bound the perturbation terms appearing in~\eqref{eq:perturbed_gradient_descent} as~\cite[Lemma 2]{Vlaski19nonconvexP1}:
\begin{align}
		{\left( \E {\|\boldsymbol{d}_{i-1}\|}^2 \right)}^2 &\le \E {\|\boldsymbol{d}_{i-1}\|}^4 \le O(\mu^4) \label{eq:d_omufourth}\\
		{\left( \E \left \{ {\|\boldsymbol{s}_{i}\|}^2 | \boldsymbol{\mathcal{F}}_{i-1} \right \}\right)}^2 &\le \E \left \{ {\|\boldsymbol{s}_{i}\|}^4 | \boldsymbol{\mathcal{F}}_{i-1} \right \} \le \sigma^4
	\end{align}
after sufficient iterations \( i \ge i_0 \). We conclude that all iterates, after sufficient iterations, approximately track the network centroid \(\w_{c, i} \), which in turn follows a perturbed gradient descent recursion, where the perturbation terms can be appropriately bounded.

We then proceeded to study the evolution of the network centroid and establish expected descent in the large gradient regime, i.e.:
\begin{align}\label{eq:descent_in_g}
  &\:\E \left \{ J(\w_{c, i}) | \w_{c, i-1} \in \mathcal{G} \right \} \notag \\
  \le&\: \E \left \{ J(\w_{c, i-1}) | \w_{c, i-1} \in \mathcal{G} \right \} - \mu^2 \frac{c_2}{\pi} + \frac{O(\mu^3)}{\pi_{i-1}^{\mathcal{G}}}
\end{align}
where the set \( \mathcal{G} \) introduced in Definition~\ref{DEF:SETS} denotes the set of points with sufficiently large gradients \( {\left \| \nabla J(w) \right \|}^2 \ge O(\mu) \).

\begin{figure*}
	\centering
  \begin{tikzpicture}[grow cyclic, align=flush center,
                      level 1/.style={level distance=3.5cm, sibling angle = 40},
                      level 2/.style={level distance=4cm, sibling angle = 30}]
                      level 3/.style={level distance=7cm, sibling angle = 30}]
    \node[rounded corners, draw=blue!60, fill=blue!20, thick]{Network centroid \\\( \w_{c, i} \) at time \( i \)} [counterclockwise from=-20]
    	child {
      node[rounded corners, draw=red!60, fill=red!20, thick] {\textbf{NOT} \( O(\mu) \)-stationary \\  \( \|\nabla J(\w_{c, i})\|^2 > O(\mu) \)} [counterclockwise from=-15]
        child {
        node[rounded corners, draw=red!60, thick, fill=red!20] {Descent in one iteration in Part I~\cite[Theorem 2]{Vlaski19nonconvexP1}: \\  \( \E \left \{ J(\w_{c, i}) - J(\w_{c, i+1}) | \w_{c, i} \in \mathcal{G} \right \} \ge O(\mu^2) \)}
        }
      }
    	child {
      node[rounded corners, draw=blue!60, fill=blue!20, thick] {\( O(\mu) \)-stationary \\  \( \|\nabla J(\w_{c, i})\|^2 \le O(\mu) \)} [counterclockwise from=-15]
        child {
          node[rounded corners, draw=green!60, thick, fill=green!20] {\( \tau \)-strict-saddle}
          child {
            node[rounded corners, draw=green!60, fill=green!20, thick]{Descent in \( i^s = O(1/(\mu \tau)) \) iterations in Theorem~\ref{TH:DESCENT_THROUGH_SADDLE_POINTS}: \\  \( \E \left \{ J(\w_{c, i}) - J(\w_{c, i+i^s}) | \w_{c, i} \in \mathcal{H} \right \} \ge O(\mu) \)}
          }
        }
        child {
          node[rounded corners, draw=blue!60, thick, fill=blue!20] { \( \lambda_{\min}\left( \nabla^2 J(\w_{c, i}  \right) > -\tau  \)  }
          child {
            node[rounded corners, draw=blue!60, fill=blue!20, thick]{\( \w_{c, i} \) is approximately second-order stationary.}
          }
        }
      }
  ;
  \end{tikzpicture}
	\caption{Classification of approximately stationary points. Theorem~\ref{TH:DESCENT_THROUGH_SADDLE_POINTS} in this work establishes descent in the green branch. The red branch is treated in Part I~\cite[Theorem 2]{Vlaski19nonconvexP1}. The two results are combined in Theorem~\ref{TH:FINAL_THEOREM} to establish the return of a second-order stationary point with high probability.}\label{fig:classification_points}
\end{figure*}
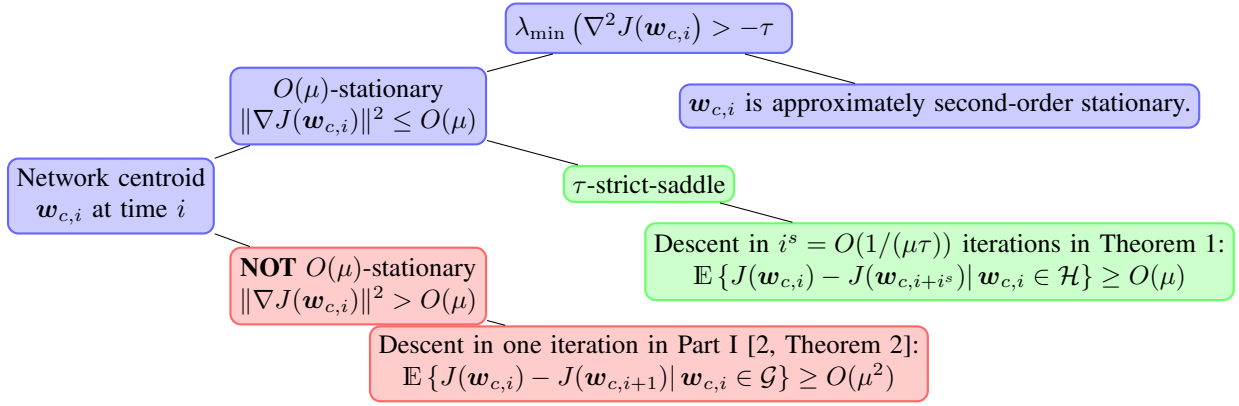

While this argument could have been continued to establish the return of approximately first-order stationary points in the complement \( \mathcal{G}^C = \mathcal{M} \cup \mathcal{H} \), our objective here is to establish the return of second-order stationary points in \( \mathcal{M} \), which is a subset of \( \mathcal{G}^C \). {This} requires the escape from strict-saddle points in \( \mathcal{H} \). In the vicinity of first-order stationary points, a single gradient step is no longer sufficient to guarantee descent, and as such it is necessary to study the cumulative effect of the gradient, as well as perturbations, over several iterations. We laid the ground work for this in Part I~\cite{Vlaski19nonconvexP1} by introducing a short-term model, which is more tractable and sufficiently accurate for a limited number of iterations. This approach has been used successfully to accurately quantify the performance of adaptive networks in convex environments~\cite{Sayed14} and establish the ability of centralized perturbed gradient descent to escape saddle-points~\cite{Ge15}. Around a first-order stationary points \( \w_{c, i^{\star}} \) at time \( i^{\star} \), the short-term model is obtained by first applying the mean-value theorem to~\eqref{eq:perturbed_gradient_descent} and obtain:
\begin{align}
  \widetilde{\w}_{i+1}^{i^{\star}} = &\: \left( I - \mu \boldsymbol{H}_{i^{\star} + i} \right) \widetilde{\w}_{i}^{i^{\star}} + \mu {\nabla J} (\w_{c, i^{\star}}) \notag \\
  &\: + \mu \boldsymbol{d}_{i^{\star}+i} + \mu \s_{i^{\star}+i+1} \label{eq:error_recursion}
\end{align}
where \( \widetilde{\w}_{i}^{i^{\star}} \) denotes the deviation from the initial point \( \w_{c, i^{\star}} \), i.e. \( \widetilde{\w}_{i}^{i^{\star}} = \w_{c, i^{\star}} - \w_{c, i^{\star}+i} \) and
\begin{equation}
  \boldsymbol{H}_{i^{\star}+i} \triangleq \int_0^1 \nabla^2 J\left( (1-t) \w_{c, i^{\star}+i} + t \w_{c, i^{\star}} \right) dt
\end{equation}
The short-term model is then obtained by replacing \( \boldsymbol{H}_{i^{\star}+i} \) by \( \nabla^2 J( \w_{c, i^{\star}}) \) and dropping the driving term \( \mu \boldsymbol{d}_{i^{\star}+i} \):
\begin{align}
  \widetilde{\w}'{}^{i^{\star}}_{i+1} =&\: \left( I - \mu \nabla^2 J( \w_{c, i^{\star}}) \right) \widetilde{\w}_{i}'{}^{i^{\star}} + \mu \nabla J(\w_{c, i^{\star}}) + \mu \s_{i^{\star}+i+1} \label{eq:long_term_recursive}
\end{align}
where again \( \widetilde{\w}'{}^{i^{\star}}_{i} \) denotes the deviation from the initialization \( \widetilde{\w}_{i}'{}^{i^{\star}} = \w_{c, i^{\star}} - \w_{c, i^{\star}+i}' \). In~\cite[Lemma 4]{Vlaski19nonconvexP1}, we established that the short-term model~\eqref{eq:long_term_recursive} is a meaningful approximation of~\eqref{eq:error_recursion} in the sense that for a limited number of iterations \( i \le \frac{T}{\mu} \), we have the following bounds:
\begin{align}
  \E \left \{ {\left \| \widetilde{\w}_{i}^{i^{\star}} \right \|}^2 | \w_{c, i^{\star}} \in \mathcal{H} \right \} &\le O(\mu) + \frac{O(\mu^2)}{\pi_{i^{\star}}^{\mathcal{H}}} \label{eq:ms_stability}\\
    \E \left \{ {\left \| \widetilde{\w}_{i}^{i^{\star}} \right \|}^3 | \w_{c, i^{\star}} \in \mathcal{H} \right \} &\le O(\mu^{3/2}) + \frac{O(\mu^3)}{{\pi_{i^{\star}}^{\mathcal{H}}}} \label{eq:mt_stability}\\
    \E \left \{ {\left \| \widetilde{\w}_{i}^{i^{\star}} \right \|}^4 | \w_{c, i^{\star}} \in \mathcal{H} \right \} &\le O(\mu^{2}) + \frac{O(\mu^{4})}{\pi_{i^{\star}}^{\mathcal{H}}} \label{eq:mf_stability}\\
    \E \left \{ {\left \| \widetilde{\w}_{i}^{i^{\star}} - \widetilde{\w}_{i}'{}^{i^{\star}} \right \|}^2 | \w_{c, i^{\star}} \in \mathcal{H} \right \} &\le O(\mu^{2}) + \frac{O(\mu^{2})}{\pi_{i^{\star}}^{\mathcal{H}}} \label{eq:model_deviation}\\
    \E \left \{ {\left \| \widetilde{\w}_{i}'{}^{i^{\star}} \right \|}^2 | \w_{c, i^{\star}} \in \mathcal{H} \right \} &\le O(\mu) + \frac{O(\mu^2)}{\pi_{i^{\star}}^{\mathcal{H}}} \label{eq:longterm_deviation}
\end{align}
We will now proceed to argue that these deviation bounds allow us to establish decent of~\eqref{eq:error_recursion} by means of studying descent of~\eqref{eq:long_term_recursive} and leverage this fact to show that the diffusion strategy will continue to descend through strict-saddle points in Theorem~\ref{TH:DESCENT_THROUGH_SADDLE_POINTS}. This result, along with the descent for large gradients established in Part I~\cite[Theorem 2]{Vlaski19nonconvexP1} will allow us to guarantee the return of an approximately second-order stationary points in Theorem~\ref{TH:FINAL_THEOREM}. The argument is summarized in Fig.~\ref{fig:classification_points}.

\section{Escape from Saddle-Points}
The deviation bounds~\eqref{eq:ms_stability}--\eqref{eq:longterm_deviation} establish that, for the first \( O(1/\mu) \) iterations following a first-order stationary points \( \w_{c, i^{\star}} \), the trajectories of the true recursion~\eqref{eq:error_recursion} the short-term model~\eqref{eq:long_term_recursive} will remain close. As a consequence, we are able to guarantee descent of \( J(\w_{c, i^{\star} + i}) \) by studying \( J(\w_{c, i^{\star}+i}')\). Note from~\eqref{eq:quadratic_upper} that
\begin{align}
  &\: J(\w_{c, i^{\star}+i}) \notag \\
  \le&\: J(\w_{c, i^{\star}+i}') + \nabla J\left( \w_{c, i^{\star}+i}' \right)^{\mathsf{\T}} \left( \w_{c, i^{\star}+i} - \w_{c, i^{\star}+i}' \right) \notag \\
  &\:+ \frac{\delta}{2} {\left \| \w_{c, i^{\star}+i} - \w_{c, i^{\star}+i}' \right \|}^2
\end{align}
Taking conditional expecation yields:
\begin{align}
  &\: \E \left \{ J(\w_{c, i^{\star}+i}) | \w_{c, i^{\star}} \in \mathcal{H} \right \} \notag \\
  \le&\: \E \left \{ J(\w_{c, i^{\star}+i}') | \w_{c, i^{\star}} \in \mathcal{H} \right \} \notag \\
  &\: + \E \left \{ \nabla J\left( \w_{c, i^{\star}+i}' \right)^{\mathsf{\T}} \left( \w_{c, i^{\star}+i} - \w_{c, i^{\star}+i}' \right) | \w_{c, i^{\star}} \in \mathcal{H} \right \} \notag \\
  &\:+ \frac{\delta}{2} \E \left \{ {\left \| \w_{c, i^{\star}+i} - \w_{c, i^{\star}+i}' \right \|}^2| \w_{c, i^{\star}} \in \mathcal{H} \right \}
\end{align}
The two terms appearing on the right-handside can be bounded as:
\ifarx \begin{align}
  &\: \E \left \{ \nabla J\left( \w_{c, i^{\star}+i}' \right)^{\mathsf{\T}} \left( \w_{c, i^{\star}+i} - \w_{c, i^{\star}+i}' \right) | \w_{c, i^{\star}} \in \mathcal{H} \right \} \notag \\
  \stackrel{(a)}{\le} &\: \sqrt{ \E \left \{ {\left \| \nabla J\left( \w_{c, i^{\star}+i}' \right) \right \|}^2 | \w_{c, i^{\star}} \in \mathcal{H} \right \} } \notag \\
  &\: \times \sqrt{ \E \left \{ {\left\| \w_{c, i^{\star}+i} - \w_{c, i^{\star}+i}' \right\|}^2 | \w_{c, i^{\star}} \in \mathcal{H} \right \} } \notag \\
  \stackrel{\eqref{eq:model_deviation}}{\le} &\: \sqrt{ O(\mu) }  \sqrt{ O(\mu^{2}) + \frac{O(\mu^{2})}{\pi_{i^{\star}}^{\mathcal{H}}} } \notag \\
  = &\: O\left(\mu^{3/2}\right) + \frac{O(\mu^{3/2})}{\sqrt{\pi_{i^{\star}}^{\mathcal{H}}}} \notag \\
  \stackrel{(b)}{\le}&\: O\left(\mu^{3/2}\right) + \frac{O(\mu^{3/2})}{{\pi_{i^{\star}}^{\mathcal{H}}}}
\end{align}
\fi
where \( (a) \) follows from Cauchy-Schwarz, \( (b) \) follows from \( \sqrt{\pi_{i^{\star}}^{\mathcal{H}}} \ge\pi_{i^{\star}}^{\mathcal{H}}  \) since \( \pi_{i^{\star}}^{\mathcal{H}} \le 1 \) so that:
\begin{align}
  &\: \E \left \{ J(\w_{c, i^{\star}+i}) | \w_{c, i^{\star}} \in \mathcal{H} \right \} \notag \\
  \le&\: \E \left \{ J(\w_{c, i^{\star}+i}') | \w_{c, i^{\star}} \in \mathcal{H} \right \} + O\left(\mu^{3/2}\right) + \frac{O(\mu^{3/2})}{{\pi_{i^{\star}}^{\mathcal{H}}}}
\end{align}
We conclude that the function value at \( \w_{c, i^{\star}+i} \) after \( i \) iterations is upper-bounded by the function evaluated at the short-term model \( \w_{c, i^{\star}+i}' \) with an additional approximation error that is bounded. We conclude that it is sufficient to study the dynamics of the short-term model, which is more tractable. Specifically, in light of the bound~\eqref{eq:cubic_upper} following from the Lipschitz-Hessian Assumption~\ref{as:lipschitz_hessians}, we have:
\begin{align}
  J(\w_{c, i^{\star}+i}') \le&\: J(\w_{c, i^{\star}}) - {\nabla J(\w_{c, i^{\star}})}^{\T}  \widetilde{\w}_{i}'{}^{i^{\star}} \notag \\
  &\: + \frac{1}{2} {\left \| \widetilde{\w}_{i}'{}^{i^{\star}} \right \|}_{\nabla^2 J(\w_{c, i^{\star}})}^2
  + \frac{\rho}{6} {\left \| \widetilde{\w}_{i}'{}^{i^{\star}} \right \|}^3 \label{eq:t_step_descent}
\end{align}

In order to establish escape from saddle-points, we need to carefully bound each term appearing on the right handside of~\eqref{eq:t_step_descent}, and to this end, we will need study the effect to the gradient noise term over several iterations. For this purpose, we introduce the following smoothness condition on the gradient noise covariance~\cite{Sayed14}:
\begin{assumption}[\textbf{Lipschitz covariances}]\label{as:lipschitz_covariance}
  The gradient noise process has a Lipschitz covariance matrix, i.e.,
  \begin{equation}
    R_{s, k}(\w_{k, i-1}) \triangleq \E \left \{ \s_{k, i}(\w_{k, i-1}) {\s_{k, i}(\w_{k, i-1})}^{\T} | \boldsymbol{\mathcal{F}}_{i-1}\right \}
  \end{equation}
  satisfies
  \begin{equation}\label{eq:lipschitz_r}
    \| R_{s, k}(x) - R_{s, k}(y) \| \le \beta_R {\| x - y \|}^{\gamma}
  \end{equation}
  for some \( \beta_R \) and \( 0 < \gamma \le 4\).\hfill\IEEEQED
\end{assumption}

\begin{definition}\label{def:aggregate_covariance}
  We define the aggregate gradient noise covariance as:
  \begin{equation}
    \mathcal{R}_{s, i}\left( \bcw_{i-1} \right) = \E \left \{ \s_i \s_i^{\T} | \boldsymbol{\mathcal{F}}_{i-1} \right \}
  \end{equation}
  where \( \s_i \triangleq \sum_{k=1}^N p_k \s_{k, i}\left( \w_{k, i-1} \right) \) denotes the aggregate gradient noise term introduced earlier in~\eqref{eq:centralized_gradient_noise}.\hfill\IEEEQED
\end{definition}
Note that in light of this definition and the assumption that the gradient noise process is conditionally uncorrelated over space as in~\eqref{eq:uncorrelated_noise}, we have:
\begin{align}
  &\: \mathcal{R}_{s, i}\left( \bcw_{i-1} \right) \notag \\
  =&\: \E \left \{ \s_i \s_i^{\T} | \boldsymbol{\mathcal{F}}_{i-1} \right\} \notag \\
  =&\: \E \left \{ \left(\sum_{k=1}^N p_k \s_{k, i}\left( \w_{k, i-1} \right)\right) {\left(\sum_{k=1}^N p_k \s_{k, i}\left( \w_{k, i-1} \right)\right)}^{\T} | \boldsymbol{\mathcal{F}}_{i-1} \right\}\notag \\
  =&\: \E \left \{ \sum_{k=1}^N p_k^2 \s_{k, i}\left( \w_{k, i-1} \right) \s_{k, i}\left( \w_{k, i-1} \right)^{\T} | \boldsymbol{\mathcal{F}}_{i-1} \right\} \notag \\
  \ifarx =&\: \sum_{k=1}^N p_k^2 \E \left \{  \s_{k, i}\left( \w_{k, i-1} \right) \s_{k, i}\left( \w_{k, i-1} \right)^{\T} | \boldsymbol{\mathcal{F}}_{i-1} \right\} \notag \\ \fi
  =&\: \sum_{k=1}^N p_k^2 R_{s, k}\left( \w_{k, i-1}  \right)\label{eq:calR_decomp}
\end{align}
so that the aggregate gradient noise covariance is a weighted combination of the individual gradient noise covariances, albeit evaluated at different iterates. In light of the smoothness assumption~\ref{as:lipschitz_covariance}, we are nevertheless able to approximate the aggregate noise covariance by one that is evaluated at the centroid.
\begin{lemma}[\textbf{Noise covariance at centroid}]\label{LEM:CENTROID_COVARIANCE}
	Under assumptions~\ref{as:strongly_connected}--\ref{as:lipschitz_covariance} and for sufficiently small step-sizes \( \mu \), we have for all \( i \) and \( w \in \mathds{R}^M \):
  \begin{align}
    &\left\|\mathcal{R}_{s, i}\left( \mathds{1} \otimes \w_{c, i-1} \right) - \mathcal{R}_{s, i}\left( \mathds{1} \otimes w \right)\right\| \le p_{\max} \beta_R \left\| \w_{c, i-1} - w \right\|^{\gamma} \label{eq:difference_r_small}\\
    &\left\|\mathcal{R}_{s, i}\left( \bcw_{c, i-1} \right) - \mathcal{R}_{s, i}\left( \bcw_{i-1} \right)\right\| \le p_{\max} \beta_R \left\| \bcw_{c, i-1} - \bcw_{i-1} \right\|^{\gamma} \label{eq:difference_r_large}
  \end{align}
\end{lemma}

\begin{IEEEproof}
	Appendix~\ref{ap:centroid_covariance}.
\end{IEEEproof}
{Note that from the bound on the aggregate gradient noise variance~\eqref{eq:gradientnoise}, we can upper bound the gradient noise covariance:
\begin{align}
  \left \| \mathcal{R}_{s, i} \left( \cw \right) \right \| = \left \| \E \s_i \s_i^{\T} \right \|\stackrel{(a)}{\le}\E \left \| \s_i \s_i^{\T} \right \| = \E \left \| \s_i \right\|^2 \stackrel{\eqref{eq:gradientnoise}}{\le} \sigma^2\label{eq:bounded_covariance}
\end{align}
where \( (a) \) follows from Jensen's inequality. In order to ensure escape from saddle-points, we introduce a similar, lower-bound condition.
}
\begin{assumption}[\textbf{Gradient noise in strict saddle-points}]\label{as:noise_in_saddle}
  Suppose \( w \) is an approximate strict-saddle points, i.e., \( w \in \mathcal{H} \) and denote the eigendecomposition of the Hessian as \( \nabla^2 J(w) = V \Lambda V^{\T} \). We introduce the decomposition:
  \begin{equation}
    {V} = \left[ \begin{array}{cc} {V}^{\ge0} & {V}^{< 0} \end{array} \right],
    \ \ {\Lambda} = \left[ \begin{array}{cc} {\Lambda}^{\ge0} & 0\\0 & {\Lambda}^{< 0} \end{array}\right]
  \end{equation}
  where \( {\Lambda}^{\ge0} \ge 0 \) and \( {\Lambda}^{< 0} < 0 \). Then, we assume that:
  \begin{equation}
    \lambda_{\min}\left({\left({V}^{< 0}\right)}^{\T} \mathcal{R}_{s}\left(\mathds{1}\otimes w \right) {V}^{< 0} \right) \ge \sigma_{\ell}^2
  \end{equation}
  for some \( \sigma_{\ell}^2 > 0 \) and all \( w \in \mathcal{H} \).\hfill\IEEEQED
\end{assumption}
Assumption~\ref{as:noise_in_saddle} is similar to the condition in~\cite{HadiDaneshmand18}, where alternating step-sizes are employed, and essentially states than for every strict-saddle point in the set \( \mathcal{H}\), there is gradient noise present along some descent direction, spanned by the eigenvectors corresponding to the negative eigenvalues of the Hessian \( \nabla^2 J(\cdot) \).

\begin{theorem}[Descent through strict saddle-points]\label{TH:DESCENT_THROUGH_SADDLE_POINTS}
  {Suppose \( \mathrm{Pr} \left \{ \w_{c, i^{\star}} \in \mathcal{H} \right \} \neq 0 \), i.e., \( \w_{c, i^{\star}} \)} is approximately stationary with significant negative eigenvalue. Then, iterating for \( i^s \) iterations after \( i^{\star} \) with
  \begin{align}
    i^{s} = {\log\left( 2 M  \frac{\sigma^2}{\sigma_{\ell}^2} + 1 \right)}{\log({1 + 2\mu\tau})} \le O\left(\frac{1}{\mu \tau} \right)
  \end{align}
  guarantees
  \begin{align}
    &\: \E \left \{ J(\w_{c, i^{\star}+i^s}) | \w_{c, i^{\star}} \in \mathcal{H} \right \} \notag \\
    \le&\: \E \left \{ J(\w_{c, i^{\star}}) | \w_{c, i^{\star}} \in \mathcal{H} \right \} - \frac{\mu}{2} M \sigma_u^2 + o(\mu) + \frac{o(\mu)}{\pi_{i^{\star}}^{\mathcal{H}}}
  \end{align}
\end{theorem}
\begin{IEEEproof}
  Appendix~\ref{AP:DESCENT_THROUGH_SADDLE_POINTS}.
\end{IEEEproof}
This result establishes that, even if \( \w_{c, i^{\star}} \) is an \( O(\mu) \)-square-stationary point and Part I~\cite[Theorem 2]{Vlaski19nonconvexP1} can no longer guarantee sufficient descent, the expected function value at the network centroid will continue to decrease, as long as the Hessian matrix has a sufficiently negative eigenvalue.

\section{Main Result}
In Part I~\cite[Theorem 2]{Vlaski19nonconvexP1}, we established a descent condition for points with large gradient norm \( \w_{c, i} \in \mathcal{G} \), while Theorem~\ref{TH:DESCENT_THROUGH_SADDLE_POINTS} guarantees descent in \( i^s \) iterations for strict-saddle points \( \w_{c, i} \in \mathcal{H} \). Together, they establish descent whenever \( \w_{c, i} \in \mathcal{G} \cup \mathcal{H} = \mathcal{M}^C \). Hence, we conclude that, as long as the cost is bounded from below, the algorithm must necessarily reach a point in \( \mathcal{M} \) after a finite amount of iterations. This intuition is formalized in the following theorem.
\begin{theorem}\label{TH:FINAL_THEOREM}
  For sufficiently small step-sizes \( \mu \), we have with probability \( 1 - \pi \), that \( \w_{c, i^o} \in \mathcal{M} \), i.e., \( \| \nabla J(\w_{c, i^o}) \|^2 \le O(\mu) \) and \( \lambda_{\min}\left( \nabla^2 J(\w_{c, i^o}) \right) \ge -\tau \) in at most \( i^o \) iterations, where
  \begin{align}
    i^o \le \frac{\left( J(w_{c, 0}) - J^o \right)}{\mu^2 c_2 \pi} i^s
  \end{align}
  and \( i^s \) denotes the escape time from Theorem~\ref{TH:DESCENT_THROUGH_SADDLE_POINTS}, i.e.,
  \begin{align}
    i^{s} = \frac{\log\left( 2 M  \frac{\sigma^2}{\sigma_{\ell}^2} + 1 \right)}{\log({1 + 2\mu\tau})} \le O\left(\frac{1}{\mu \tau} \right)
  \end{align}
\end{theorem}
\begin{IEEEproof}
  Appendix~\ref{AP:FINAL_THEOREM}.
\end{IEEEproof}
{This final result states that with probability \( 1 - \pi \), where we are free to choose the desired confidence level, the diffusion strategy~\eqref{eq:adapt}--\eqref{eq:combine} will have visited an approximately second-order stationary point after at most \( i^o \) iterations.}

\section{Simulation Results}
\begin{figure}[!t]
	\centering
	\includegraphics[width=\columnwidth]{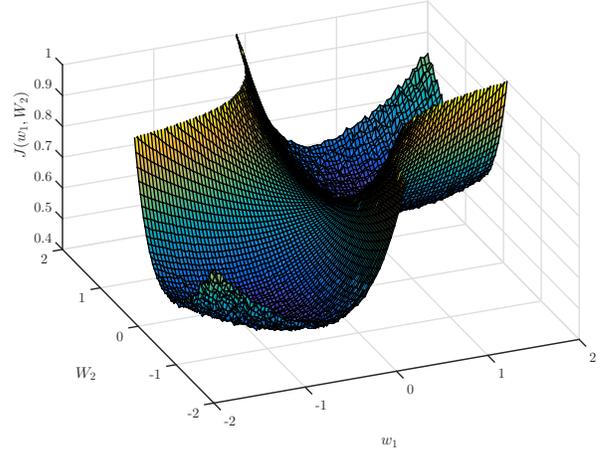}
	\caption{Cost surface of a simple neural network with \( \rho = 0.1\).}\label{fig:loss_surface}
\end{figure}

\begin{figure}[!t]
	\centering
	\includegraphics[width=\columnwidth]{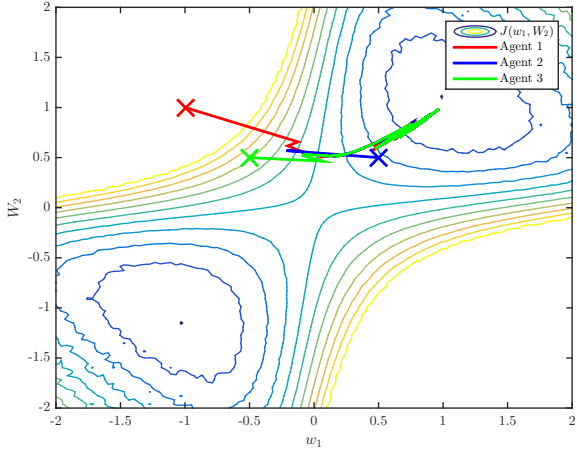}
	\caption{Agents are initialized at different points in space, but nevertheless quickly cluster. They then jointly travel away from the strict saddle-point and towards one of the local minimers.}\label{fig:different_init}
\end{figure}

\begin{figure}[!t]
	\centering
	\includegraphics[width=\columnwidth]{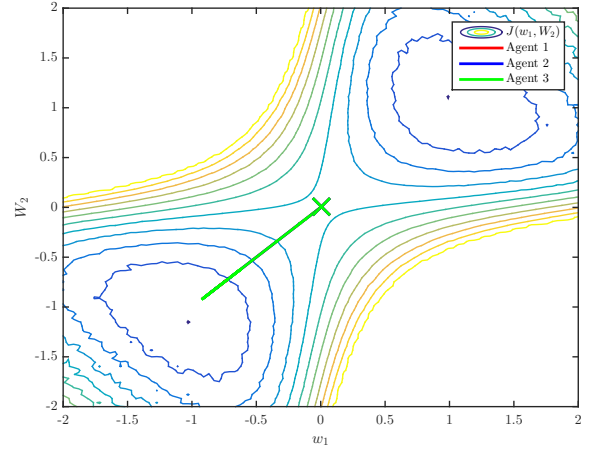}
	\caption{Agents are initialized together precisely in the strict saddle-point. The presence of the gradient perturbation allows them to jointly escape the saddle-point.}\label{fig:same_init}
\end{figure}

In this section, we consider an example that will allow us to visualize the ability of the diffusion strategy to escape saddle-points. Given a binary class label \( \boldsymbol{\gamma} \in \left \{ 0, 1 \right \} \) and feature vector \( \boldsymbol{h} \in \mathds{R}^M \), we consider a neural network with a single, linear hidden layer and a logistic activation function leading into the output layer:
\begin{equation}
  \boldsymbol{\widehat{\gamma}}\left(\h \right) \triangleq \frac{1}{1+e^{-w_1^{\T} W_2 \h}}
\end{equation}
with weights \( w_1 \in \mathds{R}^{L}, W_2 \in \mathds{R}^{L\times M}\) of appropriate dimensions. A popular risk function for training is the cross-entropy loss:
\begin{equation}\label{eq:cross_entropy}
  Q(w_1, W_2; \boldsymbol{\gamma}, \h) \triangleq - \boldsymbol{\gamma} \log(\widehat{\boldsymbol{\gamma}}) - (1-\boldsymbol{\gamma}) \log(1-\widehat{\boldsymbol{\gamma}})
\end{equation}
\ifarx Note that, the first term is non-zero, while the second term is zero if, and only if, \( \boldsymbol{\gamma} = 1 \), in which case we have:
\begin{align}
  - \boldsymbol{\gamma} \log(\widehat{\boldsymbol{\gamma}}) &= \log\left({1+e^{- w_1^{\T} W_2 \h}}\right)
\end{align}
Similarly, the second term is non-zero while the first term is zero if, and only if, \( \boldsymbol{\gamma} = 0 \), which implies:
\begin{align}
  - (1-\boldsymbol{\gamma}) \log(1-\widehat{\boldsymbol{\gamma}}) &= -\log\left(1-\frac{1}{1+e^{-w_1^{\T} W_2 \h}}\right) \notag \\
  &= -\log\left(\frac{e^{- w_1^{\T} W_2 \h}}{1+e^{- w_1^{\T} W_2 \h}}\right) \notag \\
  &= -\log\left(\frac{1}{1+e^{w_1^{\T} W_2 \h}}\right) \notag \\
  &= \log\left({1+e^{w_1^{\T} W_2 \h}}\right)
\end{align}
Letting \( \boldsymbol{\gamma}' \in \{ -1, 1 \} \) such that:
\begin{equation}
  \boldsymbol{\gamma}' \triangleq \begin{cases} -1, \ &\mathrm{if}\ \boldsymbol{\gamma} = 0 \\ 1, \ &\mathrm{if}\ \boldsymbol{\gamma} = 1. \end{cases}
\end{equation}
we can hence simplify~\eqref{eq:cross_entropy} to an equivalent logistic loss:
\begin{equation}
  Q(w_1, W_2; \boldsymbol{\gamma}', \h) = \log\left({1+e^{- \boldsymbol{\gamma}' w_1^{\T} W_2 \h}}\right)
\end{equation}
\fi
The regularized learning problem can then be formulated as:
\begin{equation}\label{eq:sample_problem}
  J(w_1, W_2) = \E Q(w_1, W_2; \boldsymbol{\gamma}', \h) + \frac{\rho}{2}\|w_1\|^2 + \frac{\rho}{2} \| W_2 \|_F^2
\end{equation}
which fits into the framework~\eqref{eq:global_problem} treated in this work. In order to be able to visualize and enumerate all stationary points of~\eqref{eq:sample_problem}, we assume in the sequel that \( M = L = 1 \) so that all involved quantities are scalar variables. We can then find:
\begin{align}
  \nabla J(w_1, W_2) &= \E \begin{pmatrix}\rho w_1 - \frac{\boldsymbol{\gamma}' W_2 \h}{e^{\boldsymbol{\gamma}' w_1 W_2 \h}} \\ \rho W_2 - \frac{\boldsymbol{\gamma}' w_1 \h}{e^{\boldsymbol{\gamma}' w_1 W_2 \h}} \end{pmatrix}
\end{align}
The cost surface is depicted in Fig.~\ref{fig:loss_surface}.
It can be observed from the figure, and analytically verified, that \( J(\cdot) \) has two local minima in the positive and negative quadrants, respectively, and a single saddle-point at \( w_1 = W_2 = 0 \). The Hessian matrix of \( J(\cdot) \) at \( w_1 = W_2 = 0 \) evaluates to:
\begin{equation}
  \nabla^2 J(0, 0) = \begin{pmatrix} \rho & - \E \frac{\boldsymbol{\gamma}'\boldsymbol{h}}{2} \\ -\E \frac{\boldsymbol{\gamma}'\boldsymbol{h}}{2} & \rho \end{pmatrix}
\end{equation}
For this example, we let \( \mathrm{Pr}\left\{ \boldsymbol{\gamma}' = -1 \right\} = \mathrm{Pr}\left\{ \boldsymbol{\gamma}' = 1 \right\} = \frac{1}{2} \) and \( \boldsymbol{h} \sim \mathcal{N} \left( \boldsymbol{\gamma}', 1 \right) \). Then, we obtain \( \E \boldsymbol{\gamma}' \h = 1\). We also let \( \rho = 0.1 \), so that:
\begin{equation}
  \nabla^2 J(0, 0)  = \begin{pmatrix} 0.1 & -0.5 \\ -0.5 & 0.1 \end{pmatrix}
\end{equation}
which has an eigenvalue at \( -0.4 \) with corresponding eigenvector \( \mathrm{col}\left\{ 1, 1 \right\} \). This implies that \( w_1 = W_2 = 0 \) is a strict saddle-point with local descent direction \( \mathrm{col}\left\{ 1, 1 \right\} \). It turns out, however that the gradient noise induced by the immediate stochastic gradient approximation \( \widehat{\nabla J}(\cdot) = \nabla Q(\cdot; \boldsymbol{\gamma}', \boldsymbol{h}) \) does not have a gradient noise component in the descent direction \( \mathrm{col}\left\{ 1, 1 \right\} \) at the strict saddle-point \( w_1 = W_2 = 0 \). Indeed, note that with probability one we have \( \nabla Q(0, 0; \boldsymbol{\gamma}', \boldsymbol{h} ) = \mathrm{col}\{ 0, 0 \} = \nabla J(0, 0) \) so that the gradient noise vanishes at \( w_1 = W_2 = 0 \). Hence, initializing all agents at \( w_1 = W_2 = 0 \) and iterating~\eqref{eq:adapt}--\eqref{eq:combine} would cause them to remain there with probability \( 1 \). This suggests that assumption~\ref{as:noise_in_saddle} is not merely a technical condition but indeed necessary. To satisfy the assumption we construct the stochastic gradient approximation as:
\begin{equation}
  \widehat{\nabla J}(w_1, W_2) \triangleq \nabla Q(w_1, W_2; \boldsymbol{\gamma}', \h) + \boldsymbol{v} \cdot \mathrm{col}\left\{ 1, 1 \right\}
\end{equation}
where \( \boldsymbol{v} \sim \mathcal{N}(0, 1) \) acts only in the direction \( \mathrm{col}\left\{ 1, 1 \right\} \) and ensures that gradient noise is present in the descent direction around the strict saddle-point at \( w_1 = W_2 = 0 \). Two realizations of the evolution are shown in Figures~\ref{fig:different_init}--\ref{fig:same_init}.

\appendices%

\section{Proof of Lemma~\ref{LEM:CENTROID_COVARIANCE}}\label{ap:centroid_covariance}
\noindent Recall that
\begin{align}
  \s_i \triangleq \sum_{k=1}^N p_k \s_{k, i}\left( \w_{k, i-1} \right)
\end{align}
and hence~\eqref{eq:calR_decomp} holds. Using the smoothness assumption on the gradient noise term~\eqref{eq:lipschitz_r}, we can write:
\begin{align}
  &\: \E \left \{ \s_i \s_i^{\T} | \boldsymbol{\mathcal{F}}_{i-1} \right\} \notag \\
  \ifarx =&\: \sum_{k=1}^N p_k^2 R_{s, k}\left( \w_{k, i-1}  \right) \notag \\ \fi
  =&\: \sum_{k=1}^N p_k^2 R_{s, k}\left( \w_{c, i-1}  \right) \notag \\
  &\: + \left( \sum_{k=1}^N p_k^2 R_{s, k}\left( \w_{k, i-1}  \right) - \sum_{k=1}^N p_k^2 R_{s, k}\left( \w_{c, i-1}  \right)\right)
\end{align}
so that:
\begin{align}
  &\: \left \| \mathcal{R}_s\left( \mathds{1} \otimes \w_{c, i-1} \right) - \mathcal{R}_s\left( \mathds{1} \otimes w \right) \right \| \notag \\
  \ifarx =&\: \left \| \sum_{k=1}^N p_k^2 R_{s, k}\left( \w_{c, i-1}  \right) - \sum_{k=1}^N p_k^2 R_{s, k}\left( w \right) \right \| \notag \\ \fi
  =&\: \left \| \sum_{k=1}^N p_k^2 \left( R_{s, k}\left( \w_{c, i-1}  \right) -  R_{s, k}\left( w \right) \right) \right \| \notag \\
  \stackrel{(a)}{\le}&\: \sum_{k=1}^N p_k \left \| p_k \left( R_{s, k}\left( \w_{c, i-1}  \right) -  R_{s, k}\left( w  \right) \right) \right \| \notag \\
  {\le}&\: p_{\max} \sum_{k=1}^N p_k\left \| R_{s, k}\left( \w_{c, i-1}  \right) -  R_{s, k}\left( w \right)  \right \| \notag \\
  \stackrel{(b)}{\le}&\: p_{\max} \beta_R \left \| \w_{c, i-1} - w \right \|^{\gamma}
\end{align}
where \( (a) \) follows from Jensen's inequality and \( (b) \) follows from the Lipschitz condition on the gradient noise covariance~\eqref{eq:lipschitz_r} and \( \sum_{k=1}^N p_k = 1 \). Similarly:
\begin{align}
  &\: \left \| R_s\left( \bcw_{i-1} \right) - R_s\left( \bcw_{c, i-1} \right) \right \| \notag \\
  \ifarx =&\: \left \| \sum_{k=1}^N p_k^2 R_{s, k}\left( \w_{k, i-1}  \right) - \sum_{k=1}^N p_k^2 R_{s, k}\left( \w_{c, i-1}  \right) \right \| \notag \\ \fi
  \ifarx =&\: \left \| \sum_{k=1}^N p_k^2 \left( R_{s, k}\left( \w_{k, i-1}  \right) -  R_{s, k}\left( \w_{c, i-1}  \right) \right) \right \| \notag \\
  \le&\: \sum_{k=1}^N p_k \left \| p_k \left( R_{s, k}\left( \w_{k, i-1}  \right) -  R_{s, k}\left( \w_{c, i-1}  \right) \right) \right \| \notag \\ \fi
  \le&\: p_{\max} \beta_R \sum_{k=1}^N p_k \left \| \w_{k, i-1} - \w_{c, i-1} \right \|^{\gamma} \notag \\
  \stackrel{(a)}{\le}&\: p_{\max} \beta_R \sum_{k=1}^N p_k \left \| \bcw_{i-1} - \bcw_{c, i-1} \right \|^{\gamma} \notag \\
  =&\: p_{\max} \beta_R \left \| \bcw_{i-1} - \bcw_{c, i-1} \right \|^{\gamma}
\end{align}
where \( (a) \) follows from the fact that \( x^{\gamma} \) is monotonically increasing in \(\gamma\) for \( x, \gamma >0 \) and:
\begin{align}
  \left \| \bcw_{i-1} - \bcw_{c, i-1} \right \|^2 =&\: \sum_{k=1}^N \left \| \w_{k, i-1} - \w_{c, i-1} \right \|^2 \notag \\
  \ge&\: \left \| \w_{\ell, i-1} - \w_{c, i-1} \right \|^2, \ \ \forall \ \ell
\end{align}

\section{Proof of Theorem~\ref{TH:DESCENT_THROUGH_SADDLE_POINTS}}\label{AP:DESCENT_THROUGH_SADDLE_POINTS}
We shall carefully bound each of the terms appearing on the righthand side of~\eqref{eq:t_step_descent}, which we repeat here again for reference:
\begin{align}
  J(\w_{c, i^{\star}+i}') \le&\: J(\w_{c, i^{\star}}) - {\nabla J(\w_{c, i^{\star}})}^{\T}  \widetilde{\w}_{i}'{}^{i^{\star}} \notag \\
  &\: + \frac{1}{2} {\left \| \widetilde{\w}_{i}'{}^{i^{\star}} \right \|}_{\nabla^2 J(\w_{c, i^{\star}})}^2  + \frac{\rho}{6} {\left \| \widetilde{\w}_{i}'{}^{i^{\star}} \right \|}^3 \label{eq:t_step_descent_app}
\end{align}
We begin by establishing a bound on the linear term in~\eqref{eq:t_step_descent_app}. Iterating the recursive relation for the short-term model~\eqref{eq:long_term_recursive} and taking expectations conditioned on \( \boldsymbol{\mathcal{F}}_{i^{\star}+i} \) yields:
\begin{align}
  &\: \E \left \{ \widetilde{\w}'{}^{i^{\star}}_{i+1} | \boldsymbol{\mathcal{F}}_{i^{\star}+i} \right \} \notag \\
  =&\: \left( I - \mu \nabla^2 J( \w_{c, i^{\star}}) \right)  \widetilde{\w}'{}^{i^{\star}}_{i}  \notag \\
  &\: + \mu \nabla J(\w_{c, i^{\star}}) + \mu \E \left \{  \s_{i^{\star} + i + 1} | \boldsymbol{\mathcal{F}}_{i^{\star}+i} \right \} \notag \\
  =&\: \left( I - \mu \nabla^2 J( \w_{c, i^{\star}}) \right)  \widetilde{\w}'{}^{i^{\star}}_{i} + \mu \nabla J(\w_{c, i^{\star}})\label{eq:interdalfsldfa}
\end{align}
where the gradient-noise term disappeared in light of
\begin{equation}
  \E \left \{  \s_{i^{\star} + i + 1} | \boldsymbol{\mathcal{F}}_{i^{\star}+i} \right \} = 0
\end{equation}
by Assumption~\ref{as:gradientnoise}. {Note that \( \boldsymbol{\mathcal{F}}_{i^{\star}+i} \) denotes the information captured in \( \w_{k, j} \) up to time \( i^{\star}+i \), while \( \boldsymbol{\mathcal{F}}_{i^{\star}} \) denotes the information available up to time \( i^{\star} \). Hence:
\begin{equation}
  \boldsymbol{\mathcal{F}}_{i^{\star}+i} = \boldsymbol{\mathcal{F}}_{i^{\star}} \cup \mathrm{filtration}\left \{ \w_{k, i^{\star}+1}, \ldots, \w_{k, i^{\star}+i} \right \}
\end{equation}
Hence, taking expectation of~\eqref{eq:interdalfsldfa} conditioned on \( \boldsymbol{\mathcal{F}}_{i^{\star}} \) removes the elements in \( \mathrm{filtration}\left \{ \w_{k, i^{\star}+1}, \ldots, \w_{k, i^{\star}+i} \right \} \) contained in \( \boldsymbol{\mathcal{F}}_{i^{\star}} \) and yields:}
\begin{align}
  \E \left \{ \widetilde{\w}'{}^{i^{\star}}_{i+1} | \boldsymbol{\mathcal{F}}_{i^{\star}} \right \} =&\: \left( I - \mu \nabla^2 J( \w_{c, i^{\star}}) \right) \E \left \{ \widetilde{\w}'{}^{i^{\star}}_{i} | \boldsymbol{\mathcal{F}}_{i^{\star}} \right \} \notag \\
  &\:+ \mu \nabla J(\w_{c, i^{\star}}) \label{eq:conditional_mean_recursion}
\end{align}
Since \( \widetilde{\w}'{}^{i^{\star}}_{0} = 0 \), iterating starting at \( i=0 \) yields:
\begin{equation}
  \E \left \{ \widetilde{\w}'{}^{i^{\star}}_{i} | \boldsymbol{\mathcal{F}}_{i^{\star}} \right \} = \mu \left( \sum_{k=1}^{i}{\left( I - \mu \nabla^2 J( \w_{c, i^{\star}}) \right)}^{k-1} \right) \nabla J(\w_{c, i^{\star}}) \label{eq:mean_deviation_saddle}
\end{equation}
This allows us to bound the linear term appearing in~\eqref{eq:t_step_descent_app} as:
\begin{align}
  &\: -\E \left\{ {\nabla J(\w_{c, i^{\star}})}^{\T}  \widetilde{\w}_{i}'{}^{i^{\star}} | \boldsymbol{\mathcal{F}}_{i^{\star}} \right \} \notag \\
  =&\: - {\nabla J(\w_{c, i^{\star}})}^{\T} \E \left \{ \widetilde{\w}_{i}'{}^{i^{\star}} | \boldsymbol{\mathcal{F}}_{i^{\star}} \right \} \notag \\
  \stackrel{\eqref{eq:mean_deviation_saddle}}{=}&\: -\mu  {\nabla J(\w_{c, i^{\star}})}^{\T} \left( \sum_{k=1}^{i} {\left( I - \mu \nabla^2 J( \w_{c, i^{\star}}) \right)}^{k-1} \right)  \nabla J(\w_{c, i^{\star}}) \notag \\
  =&\: -\mu  {\left\|\nabla J(\w_{c, i^{\star}})\right\|}^2_{\sum_{k=1}^{i} {\left( I - \mu \nabla^2 J( \w_{c, i^{\star}}) \right)}^{k-1}} \label{eq:linear_term_bound}
\end{align}
We now examine the quadratic term in~\eqref{eq:t_step_descent_app}. To this end, we introduce the eigenvalue decomposition of the Hessian around the iterate at time \( i^{\star} \):
\begin{equation}
  \nabla^2 J(\w_{c, i^{\star}}) \triangleq \boldsymbol{V}_{i^{\star}} \boldsymbol{\Lambda}_{i^{\star}} \boldsymbol{V}_{i^{\star}}^{\T}
\end{equation}
Note that both \( \boldsymbol{V}_{i^{\star}} \) and \( \boldsymbol{\Lambda}_{i^{\star}} \) inherit their randomness from \( \w_{c, i^{\star}} \). As such, they are random but become deterministic when conditioning on \( \boldsymbol{\mathcal{F}}_{i^{\star}} \). This fact will be exploited further below. To begin with, note that:
\begin{align}
  {\left \| \widetilde{\w}'{}^{i^{\star}}_{i+1} \right \|}_{\nabla^2 J(\w_{c, i^{\star}})}^2 =&\: {\left \| \widetilde{\w}'{}^{i^{\star}}_{i+1} \right \|}_{\boldsymbol{V}_{i^{\star}} \boldsymbol{\Lambda}_{i^{\star}} \boldsymbol{V}_{i^{\star}}^{\T}}^2 \notag \\
  =&\: {\left \| \boldsymbol{V}_{i^{\star}}^{\T} {\w}_{c, i^{\star}} - \boldsymbol{V}_{i^{\star}}^{\T} {\w}_{c, i^{\star}+i+1}' \right \|}_{\boldsymbol{\Lambda}_{i^{\star}}}^2 \notag \\
  =&\: {\left \| \overline{\w}'{}^{i^{\star}}_{i+1} \right \|}_{\boldsymbol{\Lambda}_{i^{\star}}}^2
\end{align}
where we introducted:
\begin{align}
  \overline{\w}'{}^{i^{\star}}_{i+1} \triangleq \boldsymbol{V}_{i^{\star}}^{\T} \widetilde{\w}'{}^{i^{\star}}_{i+1}
\end{align}
Under this transformation, recursion~\eqref{eq:long_term_recursive} is also diagonalized, yielding:
\begin{align}
  &\:\overline{\w}'{}^{i^{\star}}_{i+1} \notag \\
  \ifarx \triangleq&\: \boldsymbol{V}_{i^{\star}}^{\T}  \widetilde{\w}'{}^{i^{\star}}_{i+1}  \notag \\ \fi
  =&\: \boldsymbol{V}_{i^{\star}}^{\T} \left( I - \mu \nabla^2 J( \w_{c, i^{\star}}) \right) \boldsymbol{V}_{i^{\star}} \boldsymbol{V}_{i^{\star}}^{\T}  \widetilde{\w}'{}^{i^{\star}}_{i} \notag \\
  &\: + \mu \boldsymbol{V}_{i^{\star}}^{\T} {\nabla} J(\w_{c, i^{\star}}) + \mu \boldsymbol{V}_{i^{\star}}^{\T} \s_{i^{\star}+i+1} \notag \\
  =&\: \left( I - \mu \boldsymbol{\Lambda}_{i^{\star}} \right)   \overline{\w}'{}^{i^{\star}}_{i}  + \mu \overline{\nabla} J(\w_{c, i^{\star}}) + \mu \overline{\s}_{i^{\star}+i+1} \label{eq:long_term_transformed}
\end{align}
with \(\overline{\nabla} J(\w_{c, i^{\star}}) \triangleq \boldsymbol{V}_{i^{\star}}^{\T} {\nabla} J(\w_{c, i^{\star}}) \) and \( \overline{\s}_{i^{\star}+i+1} \triangleq \boldsymbol{V}_{i^{\star}}^{\T} \s_{i^{\star}+i+1} \). The presence of the gradient term, which is deterministic conditioned on \( \boldsymbol{\mathcal{F}}_{i^{\star}} \) complicates the analysis of the evolution. It can be removed by (conditionally) centering the random variable. Specifically, applying the same tranformation to the conditional mean recursion~\eqref{eq:conditional_mean_recursion}, and subtracting the transformed conditional mean on both sides of~\eqref{eq:long_term_transformed}, we find:
\begin{align}
  &\:\overline{\w}'{}^{i^{\star}}_{i+1} - \E \left \{ \overline{\w}'{}^{i^{\star}}_{i+1} | \boldsymbol{\mathcal{F}}_{i^{\star}} \right \} \notag \\
  =&\: \left( I - \mu \boldsymbol{\Lambda}_{i^{\star}} \right)   \left( \overline{\w}'{}^{i^{\star}}_{i} - \E \left \{ \overline{\w}'{}^{i^{\star}}_{i} | \boldsymbol{\mathcal{F}}_{i^{\star}} \right \} \right)  + \mu \overline{\s}_{i^{\star}+i+1}
\end{align}
which allows us to cancel the driving term involving the gradient. For brevity, define the (conditionally) centered random variable:
\begin{equation}
  \check{\w}'{}^{i^{\star}}_{i+1} = \overline{\w}'{}^{i^{\star}}_{i+1} - \E \left \{ \overline{\w}'{}^{i^{\star}}_{i+1} | \boldsymbol{\mathcal{F}}_{i^{\star}} \right \}
\end{equation}
so that:
\begin{align}
  \check{\w}'{}^{i^{\star}}_{i+1} = \left( I - \mu \boldsymbol{\Lambda}_{i^{\star}} \right) \check{\w}'{}^{i^{\star}}_{i}  + \mu \overline{\s}_{i^{\star}+i+1} \label{eq:centered_recursive}
\end{align}
Before proceeding, note that we can express:
\begin{align}
  &\: \E \left \{ {\left \| \check{\w}'{}^{i^{\star}}_{i} \right \|}_{\boldsymbol{\Lambda}_{i^{\star}}}^2 | \boldsymbol{\mathcal{F}}_{i^{\star}} \right \} \notag \\
  =&\: \E \left \{ {\left \| \overline{\w}'{}^{i^{\star}}_{i} - \E \left \{ \overline{\w}'{}^{i^{\star}}_{i} | \boldsymbol{\mathcal{F}}_{i^{\star}} \right \} \right \|}_{\boldsymbol{\Lambda}_{i^{\star}}}^2 | \boldsymbol{\mathcal{F}}_{i^{\star}} \right \} \notag \\
  =&\: \E \left \{ {\left \| \overline{\w}'{}^{i^{\star}}_{i} \right \|}_{\boldsymbol{\Lambda}_{i^{\star}}}^2 | \boldsymbol{\mathcal{F}}_{i^{\star}} \right \} - {\left \| \E \left \{ \overline{\w}'{}^{i^{\star}}_{i} | \boldsymbol{\mathcal{F}}_{i^{\star}} \right \} \right \|}_{\boldsymbol{\Lambda}_{i^{\star}}}^2
\end{align}
Hence, we have:
\begin{align}
  &\: \E \left \{ {\left \| \widetilde{\w}'{}^{i^{\star}}_{i} \right \|}_{\nabla^2 J(\w_{c, i^{\star}})}^2 | \boldsymbol{\mathcal{F}}_{i^{\star}} \right \} \notag \\
  =&\: \E \left \{ {\left \| \overline{\w}'{}^{i^{\star}}_{i} \right \|}_{\boldsymbol{\Lambda}_{i^{\star}}}^2 | \boldsymbol{\mathcal{F}}_{i^{\star}} \right \} \notag \\
  =&\: \E \left \{ {\left \| \check{\w}'{}^{i^{\star}}_{i} \right \|}_{\boldsymbol{\Lambda}_{i^{\star}}}^2 | \boldsymbol{\mathcal{F}}_{i^{\star}} \right \} + {\left \| \E \left \{ \overline{\w}'{}^{i^{\star}}_{i} | \boldsymbol{\mathcal{F}}_{i^{\star}} \right \} \right \|}_{\boldsymbol{\Lambda}_{i^{\star}}}^2 \label{eq:intermediate}
\end{align}
In order to make claims about \( \E \left \{ {\left \| \widetilde{\w}'{}^{i^{\star}}_{i} \right \|}_{\nabla^2 J(\w_{c, i^{\star}})}^2 | \boldsymbol{\mathcal{F}}_{i^{\star}} \right \} \) by studying \( \E \left \{ {\left \| \check{\w}'{}^{i^{\star}}_{i} \right \|}_{\boldsymbol{\Lambda}_{i^{\star}}}^2 | \boldsymbol{\mathcal{F}}_{i^{\star}} \right \} \), we need to establish a bound on \( {\left \| \E \left \{ \overline{\w}'{}^{i^{\star}}_{i} | \boldsymbol{\mathcal{F}}_{i^{\star}} \right \} \right \|}_{\boldsymbol{\Lambda}_{i^{\star}}}^2 \). We have:
\begin{align}
  &\: {\left \| \E \left \{ \overline{\w}'{}^{i^{\star}}_{i} | \boldsymbol{\mathcal{F}}_{i^{\star}} \right \} \right \|}_{\boldsymbol{\Lambda}_{i^{\star}}}^2 \notag \\
  =&\: {\left \| \E \left \{ \boldsymbol{V}_{i^{\star}}^{\T} \widetilde{\w}'{}^{i^{\star}}_{i} | \boldsymbol{\mathcal{F}}_{i^{\star}} \right \} \right \|}_{\boldsymbol{\Lambda}_{i^{\star}}}^2 \notag \\
  \stackrel{\eqref{eq:mean_deviation_saddle}}{=}&\:\mu^2 {\left \| \boldsymbol{V}_{i^{\star}}^{\T} \left( \sum_{k=1}^{i}{\left( I - \mu \nabla^2 J( \w_{c, i^{\star}}) \right)}^{k-1} \right) \nabla J(\w_{c, i^{\star}})  \right \|}_{\boldsymbol{\Lambda}_{i^{\star}}}^2 \notag \\
  =&\:\mu^2 {\left \| \left( \sum_{k=1}^{i}{\left( I - \mu \boldsymbol{\Lambda}_{i^{\star}} \right)}^{k-1} \right) \overline{\nabla} J(\w_{c, i^{\star}}) \right \|}_{\boldsymbol{\Lambda}_{i^{\star}}}^2 \notag \\
  =&\:\mu^2 {\overline{\nabla} J(\w_{c, i^{\star}}) }^{\T} \left( \sum_{k=1}^{i}{\left( I - \mu \boldsymbol{\Lambda}_{i^{\star}} \right)}^{k-1} \right) \boldsymbol{\Lambda}_{i^{\star}} \notag \\
  &\: \ \ \ \ \times \left( \sum_{k=1}^{i}{\left( I - \mu \boldsymbol{\Lambda}_{i^{\star}} \right)}^{k-1} \right) \overline{\nabla} J(\w_{c, i^{\star}})\label{eq:intermediate_some_variance}
\end{align}
{We shall order the eigenvalues of \( \nabla^2 J(\w_{c, i^{\star}}) \), such that its eigendecomposition has a block structure:
\begin{equation}\label{eq:hessian_eigendecomposition}
  \boldsymbol{V}_{i^{\star}} = \left[ \begin{array}{cc} \boldsymbol{V}_{i^{\star}}^{\ge0} & \boldsymbol{V}_{i^{\star}}^{< 0} \end{array} \right],
  \ \ \boldsymbol{\Lambda}_{i^{\star}} = \left[ \begin{array}{cc} \boldsymbol{\Lambda}_{i^{\star}}^{\ge0} & 0\\0 & \boldsymbol{\Lambda}_{i^{\star}}^{< 0} \end{array}\right]
\end{equation}
with \( \delta I \ge \boldsymbol{\Lambda}_{i^{\star}}^{\ge0} \ge 0 \) and \( \boldsymbol{\Lambda}_{i^{\star}}^{< 0} < 0 \). Note that since \( \nabla^2 J(\w_{c, i^{\star}}) \) is random, the decomposition itself is random as well. Nevertheless, it exists with probability one. We also decompose the transformed gradient vector with appropriate dimesions:
\begin{equation}
  {\overline{\nabla} J(\w_{c, i^{\star}}) } = \mathrm{col}\left \{ {\overline{\nabla} J(\w_{c, i^{\star}}) }^{\ge0}, {\overline{\nabla} J(\w_{c, i^{\star}}) }^{<0} \right \}
\end{equation}
We can then decompose~\eqref{eq:intermediate_some_variance}:
\begin{align}
  &\: {\left \| \E \left \{ \overline{\w}'{}^{i^{\star}}_{i} | \boldsymbol{\mathcal{F}}_{i^{\star}} \right \} \right \|}_{\boldsymbol{\Lambda}_{i^{\star}}}^2 \notag \\
  \ifarx {=}&\:\mu^2 {\overline{\nabla} J(\w_{c, i^{\star}}) }^{\T} \left( \sum_{k=1}^{i}{\left( I - \mu \boldsymbol{\Lambda}_{i^{\star}} \right)}^{k-1} \right) \boldsymbol{\Lambda}_{i^{\star}} \notag \\
  &\: \ \ \ \ \times \left( \sum_{k=1}^{i}{\left( I - \mu \boldsymbol{\Lambda}_{i^{\star}} \right)}^{k-1} \right) \overline{\nabla} J(\w_{c, i^{\star}}) \notag \\ \fi
  \ifarx =&\: \mu^2 {\left({\overline{\nabla} J(\w_{c, i^{\star}}) }^{\ge0}\right)}^{\T} \left( \sum_{k=1}^{i}{\left( I - \mu \boldsymbol{\Lambda}_{i^{\star}}^{\ge0} \right)}^{k-1} \right) \boldsymbol{\Lambda}_{i^{\star}}^{\ge0} \notag \\
  &\: \ \ \ \ \times \left( \sum_{k=1}^{i}{\left( I - \mu \boldsymbol{\Lambda}_{i^{\star}}^{\ge0} \right)}^{k-1} \right) \overline{\nabla} J(\w_{c, i^{\star}})^{\ge0} \notag \\
  &\:+ \mu^2 {\left({\overline{\nabla} J(\w_{c, i^{\star}}) }^{<0}\right)}^{\T} \left( \sum_{k=1}^{i}{\left( I - \mu \boldsymbol{\Lambda}_{i^{\star}}^{<0} \right)}^{k-1} \right) \boldsymbol{\Lambda}_{i^{\star}}^{<0} \notag \\
  &\: \ \ \ \ \times \left( \sum_{k=1}^{i}{\left( I - \mu \boldsymbol{\Lambda}_{i^{\star}}^{<0} \right)}^{k-1} \right) \overline{\nabla} J(\w_{c, i^{\star}})^{<0} \notag \\ \fi
  \stackrel{(a)}{\le}&\: \mu^2 {\left({\overline{\nabla} J(\w_{c, i^{\star}}) }^{\ge0}\right)}^{\T} \left( \sum_{k=1}^{i}{\left( I - \mu \boldsymbol{\Lambda}_{i^{\star}}^{\ge0} \right)}^{k-1} \right) \boldsymbol{\Lambda}_{i^{\star}}^{\ge0} \notag \\
  &\: \ \ \ \ \times \left( \sum_{k=1}^{i}{\left( I - \mu \boldsymbol{\Lambda}_{i^{\star}}^{\ge0} \right)}^{k-1} \right) \overline{\nabla} J(\w_{c, i^{\star}})^{\ge0} \notag \\
  \stackrel{(b)}{\le}&\: \mu^2 {\left({\overline{\nabla} J(\w_{c, i^{\star}}) }^{\ge0}\right)}^{\T} \left( \sum_{k=1}^{\infty}{\left( I - \mu \boldsymbol{\Lambda}_{i^{\star}}^{\ge0} \right)}^{k-1} \right) \boldsymbol{\Lambda}_{i^{\star}}^{\ge0} \notag \\
  &\: \ \ \ \ \times \left( \sum_{k=1}^{i}{\left( I - \mu \boldsymbol{\Lambda}_{i^{\star}}^{\ge0} \right)}^{k-1} \right) \overline{\nabla} J(\w_{c, i^{\star}})^{\ge0} \notag \\
  \stackrel{(c)}{=}&\: \mu^2 {\left({\overline{\nabla} J(\w_{c, i^{\star}}) }^{\ge0}\right)}^{\T} \left( \mu \boldsymbol{\Lambda}_{i^{\star}}^{\ge0} \right)^{-1} \boldsymbol{\Lambda}_{i^{\star}}^{\ge0} \notag \\
  &\: \ \ \ \ \times \left( \sum_{k=1}^{i}{\left( I - \mu \boldsymbol{\Lambda}_{i^{\star}}^{\ge0} \right)}^{k-1} \right) \overline{\nabla} J(\w_{c, i^{\star}})^{\ge0} \notag \\
  {=}&\: \mu{\left({\overline{\nabla} J(\w_{c, i^{\star}}) }^{\ge0}\right)}^{\T}  \left( \sum_{k=1}^{i}{\left( I - \mu \boldsymbol{\Lambda}_{i^{\star}}^{\ge0} \right)}^{k-1} \right) \overline{\nabla} J(\w_{c, i^{\star}})^{\ge0} \notag \\
  \stackrel{(d)}{\le}&\: \mu{\left({\overline{\nabla} J(\w_{c, i^{\star}}) }^{\ge0}\right)}^{\T}  \left( \sum_{k=1}^{i}{\left( I - \mu \boldsymbol{\Lambda}_{i^{\star}}^{\ge0} \right)}^{k-1} \right) \overline{\nabla} J(\w_{c, i^{\star}})^{\ge0} \notag \\
  &\:+ \mu{\left({\overline{\nabla} J(\w_{c, i^{\star}}) }^{<0}\right)}^{\T}  \left( \sum_{k=1}^{i}{\left( I - \mu \boldsymbol{\Lambda}_{i^{\star}}^{<0} \right)}^{k-1} \right) \overline{\nabla} J(\w_{c, i^{\star}})^{<0} \notag \\
  {\le}&\: \mu{{\overline{\nabla} J(\w_{c, i^{\star}}) }}^{\T}  \left( \sum_{k=1}^{i}{\left( I - \mu \boldsymbol{\Lambda}_{i^{\star}}\right)}^{k-1} \right) \overline{\nabla} J(\w_{c, i^{\star}}) \notag \\
  {=}&\: \mu{\left \|{\overline{\nabla} J(\w_{c, i^{\star}}) }\right \|}^{2}_{\sum_{k=1}^{i}{\left( I - \mu \boldsymbol{\Lambda}_{i^{\star}}\right)}^{k-1}}\label{eq:centered_error}
\end{align}
where \( (a) \) follows from \( \boldsymbol{\Lambda}_{i^{\star}}^{<0} < 0 \), \( (b) \) follows from:
\begin{equation}
    \sum_{k=1}^{k}{\left( I - \mu \boldsymbol{\Lambda}_{i^{\star}}^{\ge0} \right)}^{k-1}  \le  \sum_{k=1}^{\infty}{\left( I - \mu \boldsymbol{\Lambda}_{i^{\star}}^{\ge0} \right)}^{k-1}
\end{equation}
for \( \mu < \frac{1}{\delta} \). Step \( (c) \) follows from the formula for the geometric matrix series, and \( (d) \) follows from:
\begin{equation}
 \mu{\left({\overline{\nabla} J(\w_{c, i^{\star}}) }^{\ge0}\right)}^{\T}  \left( \sum_{k=1}^{i}{\left( I - \mu \boldsymbol{\Lambda}_{i^{\star}}^{\ge0} \right)}^{k-1} \right) \overline{\nabla} J(\w_{c, i^{\star}})^{\ge0} \ge 0
\end{equation}}
Comparing~\eqref{eq:centered_error} to~\eqref{eq:linear_term_bound}, we find that we can bound:
\begin{align}
  -\E \left\{ {\nabla J(\w_{c, i^{\star}})}^{\T}  \widetilde{\w}_{i}'{}^{i^{\star}} | \boldsymbol{\mathcal{F}}_{i^{\star}} \right \} + {\left \| \E \left \{ \overline{\w}'{}^{i^{\star}}_{i} | \boldsymbol{\mathcal{F}}_{i^{\star}} \right \} \right \|}_{\boldsymbol{\Lambda}_{i^{\star}}}^2 \le 0
\end{align}
To recap, we can simplify~\eqref{eq:t_step_descent_app} as:
\begin{align}
  &\: \E \left \{ J(\w_{c, i^{\star}+i}') | \boldsymbol{\mathcal{F}}_{i^{\star}} \right \} \notag \\
  \le&\: J(\w_{c, i^{\star}}) + \frac{1}{2} \E \left \{ {\left \| \check{\w}'{}^{i^{\star}}_{i} \right \|}_{\boldsymbol{\Lambda}_{i^{\star}}}^2 | \boldsymbol{\mathcal{F}}_{i^{\star}} \right \} + \frac{\rho}{6} \E \left \{ {\left \| \widetilde{\w}_{i}'{}^{i^{\star}} \right \|}^3 | \boldsymbol{\mathcal{F}}_{i^{\star}} \right \}\label{eq:t_step_descent_simplified}
\end{align}
We proceed with the now simplified quadratic term.
We square both sides of~\eqref{eq:centered_recursive} under an arbitrary diagonal weighting matrix \( \boldsymbol{\Sigma}_i \), deterministic conditioned on \( \w_{c, i^{\star}} \) and \( \w_{c, i^{\star}+i} \), to obtain:
\begin{align}
  &{\left \| \check{\w}'{}^{i^{\star}}_{i+1} \right \|}_{\boldsymbol{\Sigma}_i}^2 \notag \\
  =&\: {\left \| \left( I - \mu \boldsymbol{\Lambda}_{i^{\star}} \right)  \check{\w}'{}^{i^{\star}}_{i}  + \mu \overline{\s}_{i^{\star}+i+1}  \right \|}_{\boldsymbol{\Sigma}_{i}}^2 \notag \\
  =&\: {\left \| \left( I - \mu \boldsymbol{\Lambda}_{i^{\star}} \right)  \check{\w}'{}^{i^{\star}}_{i}  \right \|}_{\boldsymbol{\Sigma}_{i}}^2 + \mu^2  {\left \| \overline{\s}_{i^{\star}+i+1}  \right \|}_{\boldsymbol{\Sigma}_{i}}^2 \notag \\
  &+ 2 \mu { \check{\w}'{}^{i^{\star}}_{i} }^{\T} \left( I - \mu \boldsymbol{\Lambda}_{i^{\star}} \right) \boldsymbol{\Sigma}_{i} \overline{\s}_{i^{\star}+i+1}
\end{align}
Note that upon conditioning on \( \boldsymbol{\mathcal{F}}_{i^{\star}+i} \), all elements of the cross-term, aside from \( \overline{\s}_{i^{\star}+i+1} \), become deterministic, and as such the term disappears when taking expecations. We obtain:
\begin{align}
  &\: \E \left \{ {\left \| \check{\w}'{}^{i^{\star}}_{i+1} \right \|}_{\boldsymbol{\Sigma}_{i}}^2 | \boldsymbol{\mathcal{F}}_{i^{\star}+i} \right \} \notag \\
  =&\: {\left \| \left( I - \mu \boldsymbol{\Lambda}_{i^{\star}} \right)  \check{\w}'{}^{i^{\star}}_{i}  \right \|}_{\boldsymbol{\Sigma}_{i}}^2 + \mu^2 \E \left \{ {\left \| \overline{\s}_{i^{\star}+i+1}  \right \|}_{\boldsymbol{\Sigma}_{i}}^2 | \boldsymbol{\mathcal{F}}_{i^{\star}+i} \right \} \notag \\
  \ifarx =&\: {\left \| \check{\w}'{}^{i^{\star}}_{i} \right \|}_{\boldsymbol{\Sigma}_{i} - 2 \mu \boldsymbol{\Lambda}_{i^{\star}} \boldsymbol{\Sigma}_{i} + \mu^2 \boldsymbol{\Lambda}_{i^{\star}} \boldsymbol{\Sigma}_{i} \boldsymbol{\Lambda}_{i^{\star}}}^2 \notag \\
  &\: + \mu^2 \mathrm{Tr}\left( \boldsymbol{V}_{i^{\star}} \boldsymbol{\Sigma}_i \boldsymbol{V}_{i^{\star}}^{\T} \mathcal{R}_{s}\left( \bcw_{i^{\star}+i} \right) \right) \notag \\ \fi
  =&\: {\left \| \check{\w}'{}^{i^{\star}}_{i} \right \|}_{\boldsymbol{\Sigma}_{i} - 2 \mu \boldsymbol{\Lambda}_{i^{\star}} \boldsymbol{\Sigma}_{i} }^2 + \mu^2 \mathrm{Tr}\left( \boldsymbol{V}_{i^{\star}} \boldsymbol{\Sigma}_i \boldsymbol{V}_{i^{\star}}^{\T} \mathcal{R}_{s}\left( \bcw_{c, i^{\star}} \right) \right) \notag \\
  &\: + \mu^2 \mathrm{Tr}\left( \boldsymbol{V}_{i^{\star}} \boldsymbol{\Sigma}_i \boldsymbol{V}_{i^{\star}}^{\T} \left( \mathcal{R}_{s}\left( \bcw_{i^{\star}+i} \right) - \mathcal{R}_{s}\left( \bcw_{c, i^{\star}} \right)\right) \right) \notag \\
  &\: + \mu^2 {\left \| \check{\w}'{}^{i^{\star}}_{i} \right \|}_{\boldsymbol{\Lambda}_{i^{\star}} \boldsymbol{\Sigma}_{i} \boldsymbol{\Lambda}_{i^{\star}}}^2
\end{align}
We proceed to bound the last two terms. First, we have:
\begin{align}
  &\:\mathrm{Tr}\left( \boldsymbol{V}_{i^{\star}} \boldsymbol{\Sigma}_i \boldsymbol{V}_{i^{\star}}^{\T} \left( \mathcal{R}_{s}\left( \bcw_{i^{\star}+i} \right) - \mathcal{R}_{s}\left( \bcw_{c, i^{\star}} \right)\right) \right) \notag \\
  \overset{(a)}{\le}&\: \left \| \boldsymbol{V}_{i^{\star}} \boldsymbol{\Sigma}_i \boldsymbol{V}_{i^{\star}}^{\T} \right \| \left \| \mathcal{R}_{s}\left( \bcw_{i^{\star}+i} \right) - \mathcal{R}_{s}\left( \bcw_{c, i^{\star}} \right) \right \| \notag \\
  \ifarx {\le}&\: \left \| \boldsymbol{V}_{i^{\star}} \boldsymbol{\Sigma}_i \boldsymbol{V}_{i^{\star}}^{\T} \right \| \| \mathcal{R}_{s}\left( \bcw_{i^{\star}+i} \right) -  \mathcal{R}_{s}\left( \bcw_{c, i^{\star}+i} \right) \notag \\
  &\: \ \ \ \ \ \ \ \ \ \  \ \ \ \ \ \ \  +  \mathcal{R}_{s}\left( \bcw_{c, i^{\star}+i} \right)  - \mathcal{R}_{s}\left( \bcw_{c, i^{\star}} \right)  \| \notag \\ \fi
  {\le}&\: \left \| \boldsymbol{V}_{i^{\star}} \boldsymbol{\Sigma}_i \boldsymbol{V}_{i^{\star}}^{\T} \right \| \left \| \mathcal{R}_{s}\left( \bcw_{i^{\star}+i} \right) -  \mathcal{R}_{s}\left( \bcw_{c, i^{\star}+i} \right) \right \| \notag \\
  &\: + \left \| \boldsymbol{V}_{i^{\star}} \boldsymbol{\Sigma}_i \boldsymbol{V}_{i^{\star}}^{\T} \right \| \left \| \mathcal{R}_{s}\left( \bcw_{c, i^{\star}+i} \right) -  \mathcal{R}_{s}\left( \bcw_{c, i^{\star}} \right) \right \| \notag \\
  \overset{(b)}{\le}&\: \rho \left( \boldsymbol{\Sigma}_i \right) \beta_R p_{\max} \left( {\left \| \w_{c, i^{\star}+i} - \w_{c, i^{\star}} \right \|}^{\gamma} + {\left \| \bcw_{c, i^{\star}+i} - \bcw_{i^{\star}+i} \right \|}^{\gamma} \right) \notag \\
  {=}&\: \rho \left( \boldsymbol{\Sigma}_i \right) \beta_R p_{\max}\left( {\left \| \widetilde{\w}^{{i}^{\star}}_{i}\right \|}^{\gamma} + \left\|\bcw_{c, i^{\star}+i} - \bcw_{i^{\star}+i}\right\|^{\gamma} \right)
\end{align}
where \( (a) \) follows from {Cauchy-Schwarz, since \( \mathrm{Tr}(A^{\T} B) \) is an inner product over the space of symmetric matricess, and hence,} \( |\mathrm{Tr}(A^{\T} B)| \le \|A \|\|B\| \), and \( (b) \) follows from Lemma~\eqref{LEM:CENTROID_COVARIANCE}. For the second term, we have:
\begin{align}
  {\left \| \check{\w}'{}^{i^{\star}}_{i} \right \|}_{\boldsymbol{\Lambda}_{i^{\star}} \boldsymbol{\Sigma}_{i} \boldsymbol{\Lambda}_{i^{\star}}}^2 &\le \rho\left( \boldsymbol{\Lambda}_{i^{\star}} \boldsymbol{\Sigma}_{i} \boldsymbol{\Lambda}_{i^{\star}} \right) {\left \| \check{\w}'{}^{i^{\star}}_{i} \right \|}^2 \notag \\
  &\le \delta^2 \rho\left( \boldsymbol{\Sigma}_{i} \right) {\left \| \check{\w}'{}^{i^{\star}}_{i} \right \|}^2
\end{align}
We conclude that
\begin{align}
  &\: \E \left \{ {\left \| \check{\w}'{}^{i^{\star}}_{i+1} \right \|}_{\boldsymbol{\Sigma}_{i}}^2 | \boldsymbol{\mathcal{F}}_{i^{\star}} \right \} \notag \\
  =&\: \E \left \{ {\left \| \check{\w}'{}^{i^{\star}}_{i} \right \|}_{\boldsymbol{\Sigma}_{i} - 2 \mu \boldsymbol{\Lambda}_{i^{\star}} \boldsymbol{\Sigma}_{i} }^2 | \boldsymbol{\mathcal{F}}_{i^{\star}} \right \} \notag \\
  &\: + \mu^2 \mathrm{Tr}\left( \boldsymbol{V}_{i^{\star}} \boldsymbol{\Sigma}_i \boldsymbol{V}_{i^{\star}}^{\T} \mathcal{R}_{s}\left( \bcw_{c, i^{\star}} \right) \right) + \mu^2 \rho\left( \boldsymbol{\Sigma}_i \right) \E \left \{ \boldsymbol{q}_{i^{\star}+i} | \boldsymbol{\mathcal{F}}_{i^{\star}} \right \}
\end{align}
where
\begin{align}
  \boldsymbol{q}_{i^{\star}+i} \triangleq \beta_R p_{\max}\left( {\left \| \widetilde{\w}^{{i}^{\star}}_{i}\right \|}^{\gamma} + \left\|\bcw_{c, i^{\star}+i} - \bcw_{i^{\star}+i}\right\|^{\gamma} \right) + \delta^2 {\left \| \check{\w}'{}^{i^{\star}}_{i} \right \|}^2\label{eq:perturbation_definition}
\end{align}
For brevity, we define
\begin{align}
  \boldsymbol{D} &\triangleq I - 2 \mu \boldsymbol{\Lambda}_{i^{\star}} \\
  \boldsymbol{Y} &\triangleq \boldsymbol{V}_{i^{\star}}^{\T} \mathcal{R}_{s}\left( \bcw_{c, i^{\star}} \right) \boldsymbol{V}_{i^{\star}}
\end{align}
With these substitutions we obtain:
\begin{align}
  &\: \E \left \{ {\left \| \check{\w}'{}^{i^{\star}}_{i+1} \right \|}_{\boldsymbol{\Sigma}_{i}}^2 | \boldsymbol{\mathcal{F}}_{i^{\star}} \right \} \notag \\
  =&\: \E \left \{ {\left \| \check{\w}'{}^{i^{\star}}_{i} \right \|}_{\boldsymbol{D} \boldsymbol{\Sigma}_i }^2  | \boldsymbol{\mathcal{F}}_{i^{\star}} \right \} + \mu^2 \mathrm{Tr}\left( \boldsymbol{\Sigma}_i \boldsymbol{Y} \right) \notag \\
  &\: + \mu^2 \rho \left( \boldsymbol{\Sigma}_i \right) \E \left \{ \boldsymbol{q}_{i^{\star}+i} | \boldsymbol{\mathcal{F}}_{i^{\star}} \right \}
\end{align}
At \( i = 0 \), we have:
\begin{equation}
  \check{\w}'{}^{i^{\star}}_{0} = \overline{\w}'{}^{i^{\star}}_{0} - \E \left \{ \overline{\w}'{}^{i^{\star}}_{0} | \boldsymbol{\mathcal{F}}_{i^{\star}} \right \} = 0 - 0 = 0
\end{equation}
Letting \( \boldsymbol{\Sigma}_i = \boldsymbol{\Lambda}_{i^{\star}} \boldsymbol{D}^i \), we can iterate to obtain:
\begin{align}
  &\: \E \left \{ {\left \| \check{\w}'{}^{i^{\star}}_{i+1} \right \|}_{\boldsymbol{\Lambda}_{i^{\star}}}^2 | \boldsymbol{\mathcal{F}}_{i^{\star}} \right \} \notag \\
  \ifarx =&\: \mu^2 \sum_{n=0}^i \mathrm{Tr}\left( \boldsymbol{\Lambda}_{i^{\star}} \boldsymbol{D}^n \boldsymbol{Y} \right) \notag \\
  &\:+ \mu^2 \sum_{n=0}^i  \rho\left( \boldsymbol{\Lambda}_{i^{\star}} \boldsymbol{D}^n \right) \cdot \E \left \{ \boldsymbol{q}_{i^{\star}+n} | \boldsymbol{\mathcal{F}}_{i^{\star}} \right \} \notag \\ \fi
  =&\: \mu^2 \mathrm{Tr}\left( \boldsymbol{\Lambda}_{i^{\star}} \left( \sum_{n=0}^i \boldsymbol{D}^n \right) \boldsymbol{Y} \right) \notag \\
  &\:+ \mu^2 \sum_{n=0}^i  \rho\left( \boldsymbol{\Lambda}_{i^{\star}} \boldsymbol{D}^n \right) \cdot \E \left \{ \boldsymbol{q}_{i^{\star}+n} | \boldsymbol{\mathcal{F}}_{i^{\star}} \right \}\label{eq:centered_error_recursion}
\end{align}
since \( \overline{\w}_{c, i^{\star}+i+1}' = \overline{\w}_{c, i^{\star}} \) at \( i = 0 \).  Our objective is to show that the first term on the right-hand side yields sufficient descent (i.e., will be sufficiently negative), while the second term is small enough to be negligible. To this end, we again make use of the structured eigendecomposition~\eqref{eq:hessian_eigendecomposition}. We have:
\ifarx \begin{align}
  &\: \mu^2 \mathrm{Tr}\left( \boldsymbol{\Lambda}_{i^{\star}} \left( \sum_{n=0}^i \boldsymbol{D}^n \right) \boldsymbol{V}_{i^{\star}}^{\T} \mathcal{R}_{s}\left( \bcw_{c, i^{\star}} \right) \boldsymbol{V}_{i^{\star}} \right) \notag \\
  \stackrel{(a)}{=}&\: \mu^2 \mathrm{Tr}\Bigg( \boldsymbol{\Lambda}_{i^{\star}}^{\ge0} \left( \sum_{n=0}^i {\left( I - 2 \mu \boldsymbol{\Lambda}_{i^{\star}}^{\ge0} \right)}^n \right) \notag \\
  &\: \ \ \ \ \ \ \ \ \times  {\left(\boldsymbol{V}_{i^{\star}}^{\ge0}\right)}^{\T} \mathcal{R}_{s}\left( \bcw_{c, i^{\star}} \right) \boldsymbol{V}_{i^{\star}}^{\ge0} \Bigg) \notag \\
  &\:+ \mu^2 \mathrm{Tr}\Bigg( \boldsymbol{\Lambda}_{i^{\star}}^{< 0} \left( \sum_{n=0}^i {\left( I - 2 \mu \boldsymbol{\Lambda}_{i^{\star}}^{< 0} \right)}^n \right) \notag \\
  &\: \ \ \ \ \ \ \ \ \ \ \ \ \times  {\left(\boldsymbol{V}_{i^{\star}}^{< 0}\right)}^{\T} \mathcal{R}_{s}\left( \bcw_{c, i^{\star}} \right) \boldsymbol{V}_{i^{\star}}^{< 0} \Bigg) \notag \\
  \stackrel{(b)}{=}&\: \mu^2 \mathrm{Tr}\Bigg( \boldsymbol{\Lambda}_{i^{\star}}^{\ge0} \left( \sum_{n=0}^i {\left( I - 2 \mu \boldsymbol{\Lambda}_{i^{\star}}^{\ge0} \right)}^n \right) \notag \\
  &\: \ \ \ \ \ \ \ \  \times {\left(\boldsymbol{V}_{i^{\star}}^{\ge0}\right)}^{\T} \mathcal{R}_{s}\left( \bcw_{c, i^{\star}} \right) \boldsymbol{V}_{i^{\star}}^{\ge0} \Bigg) \notag \\
  &\:- \mu^2 \mathrm{Tr}\Bigg( \left(-\boldsymbol{\Lambda}_{i^{\star}}^{< 0}\right) \left( \sum_{n=0}^i {\left( I - 2 \mu \boldsymbol{\Lambda}_{i^{\star}}^{< 0} \right)}^n \right) \notag \\
  &\: \ \ \ \ \  \  \ \ \ \ \ \times {\left(\boldsymbol{V}_{i^{\star}}^{< 0}\right)}^{\T} \mathcal{R}_{s}\left( \bcw_{c, i^{\star}} \right) \boldsymbol{V}_{i^{\star}}^{< 0} \Bigg) \notag \\
  \stackrel{(c)}{\le}&\: \mu^2  \mathrm{Tr}\left( \boldsymbol{\Lambda}_{i^{\star}}^{\ge0} \left( \sum_{n=0}^i {\left( I - 2 \mu \boldsymbol{\Lambda}_{i^{\star}}^{\ge0} \right)}^n \right) \right)  \notag \\
  &\:\times \lambda_{\max}\left({\left(\boldsymbol{V}_{i^{\star}}^{\ge0}\right)}^{\T} \mathcal{R}_{s}\left( \bcw_{c, i^{\star}} \right) \boldsymbol{V}_{i^{\star}}^{\ge0} \right) \notag \\
  &\:- \mu^2 \mathrm{Tr}\left( \left(-\boldsymbol{\Lambda}_{i^{\star}}^{< 0}\right)  \left( \sum_{n=0}^i {\left( I - 2 \mu \boldsymbol{\Lambda}_{i^{\star}}^{< 0} \right)}^n \right)  \right) \notag \\
  &\:\times \lambda_{\min}\left({\left(\boldsymbol{V}_{i^{\star}}^{< 0}\right)}^{\T} \mathcal{R}_{s}\left( \bcw_{c, i^{\star}} \right) \boldsymbol{V}_{i^{\star}}^{< 0} \right) \notag \\
  \stackrel{(d)}{\le}&\: \mu^2  \mathrm{Tr}\left( \boldsymbol{\Lambda}_{i^{\star}}^{\ge0} \left( \sum_{n=0}^i {\left( I - 2 \mu \boldsymbol{\Lambda}_{i^{\star}}^{\ge0} \right)}^n \right) \right) \sigma_u^2 \notag \\
  &\:- \mu^2 \mathrm{Tr}\left( \left(-\boldsymbol{\Lambda}_{i^{\star}}^{< 0}\right)  \left( \sum_{n=0}^i {\left( I - 2 \mu \boldsymbol{\Lambda}_{i^{\star}}^{< 0} \right)}^n \right)  \right) \sigma_{\ell}^2 \label{eq:previous_term_5646346}
\end{align}
where in \( (a) \) we decomposed the trace since \( \boldsymbol{\Lambda}_{i^{\star}} \left( \sum_{n=0}^i \boldsymbol{D}^n \right) \) is a diagonal matrix, \( (b) \) applies \( - \left( - \boldsymbol{\Lambda}_{i^{\star}}^{< 0} \right) = \boldsymbol{\Lambda}_{i^{\star}}^{< 0} \).
\fi%
Step \( (b) \) follows from \( \mathrm{Tr}(A) \lambda_{\min}(B) \le \mathrm{Tr}(AB) \le \mathrm{Tr}(A) \lambda_{\max}(B) \) which holds for \( A = A^{\T}, B = B^{\T} \ge 0 \), and \( (c) \) follows from the bounded covariance property~\eqref{eq:bounded_covariance} and Assumption~\ref{as:noise_in_saddle}. For the positive term, we have:
\begin{align}
  &\: \mu^2 \mathrm{Tr}\left( \boldsymbol{\Lambda}_{i^{\star}}^{\ge0} \left( \sum_{n=0}^i {\left( I - 2 \mu \boldsymbol{\Lambda}_{i^{\star}}^{\ge0} \right)}^n \right) \right) \sigma_u^2 \notag \\
  \stackrel{(a)}{\le}&\: \mu^2  \mathrm{Tr}\left( \boldsymbol{\Lambda}_{i^{\star}}^{\ge0} \left( \sum_{n=0}^{\infty} {\left( I - 2 \mu \boldsymbol{\Lambda}_{i^{\star}}^{\ge0} \right)}^n \right) \right) \sigma_u^2  \notag \\
  \stackrel{(b)}{\le}&\: \mu^2  \mathrm{Tr}\left( \boldsymbol{\Lambda}_{i^{\star}}^{\ge0} {\left( 2 \mu \boldsymbol{\Lambda}_{i^{\star}}^{\ge0} \right)}^{-1} \right) \sigma_u^2 \stackrel{(c)}{\le}\: \frac{\mu}{2} M \sigma_u^2
\end{align}
where \( (a) \) follows since \( I - 2 \mu \boldsymbol{\Lambda}_{i^{\star}}^{\ge0} \) is elementwise non-negative for \( \mu \le \frac{2}{\delta} \), \( (b) \) follows from \( \sum_{n=0}^{\infty} A^n = {\left( I - A \right)}^{-1} \) and \( (c) \) follows since \( \nabla^2 J(\w_{c, i^{\star}}) \) is of dimension \( M \).

For the negative term, we have under expectation conditioned on \( \w_{c, i^{\star}} \in \mathcal{H} \):
\begin{align}
  &\: \E \Bigg \{ \mathrm{Tr}\left( \left(-\boldsymbol{\Lambda}_{i^{\star}}^{< 0}\right)  \left( \sum_{n=0}^i {\left( I - 2 \mu \boldsymbol{\Lambda}_{i^{\star}}^{< 0} \right)}^n \right)  \right) \sigma_{\ell}^2 \Bigg | \w_{c, i^{\star}} \in \mathcal{H} \Bigg \} \notag \\
  \stackrel{(a)}{\ge}&\: \E \left \{ \tau \left( \sum_{n=0}^{i} {\left( 1 + 2 \mu \tau \right)}^n \right) \sigma_{\ell}^2 \Bigg | \w_{c, i^{\star}} \in \mathcal{H} \right \} \notag \\
  \stackrel{(b)}=&\: \tau \left( \sum_{n=0}^{i} {\left( 1 + 2 \mu \tau \right)}^n \right) \sigma_{\ell}^2 \stackrel{(c)}{=}\:  \tau \frac{1 - {\left( 1 + 2\mu\tau \right)}^{i+1}}{1 - (1 + 2 \mu \tau)} \sigma_{\ell}^2 \notag \\
  {=}&\: \frac{1}{2\mu} \left({\left( 1 + 2\mu\tau \right)}^{i+1} - 1 \right) \sigma_{\ell}^2
\end{align}
Step \( (a) \) makes use of the fact that \( \left(-\boldsymbol{\Lambda}_{i^{\star}}^{< 0}\right)  \left( \sum_{n=0}^i {\left( I - 2 \mu \boldsymbol{\Lambda}_{i^{\star}}^{< 0} \right)}^n \right) \) is a diagonal matrix, where all elements are non-negative. Hence, its trace can be bounded by any of its diagonal elements:
\begin{align}
  &\: \mathrm{Tr}\left( \left(-\boldsymbol{\Lambda}_{i^{\star}}^{< 0}\right)  \left( \sum_{n=0}^i {\left( I - 2 \mu \boldsymbol{\Lambda}_{i^{\star}}^{< 0} \right)}^n \right)  \right) \notag \\
  \stackrel{\eqref{eq:define_h}}{\ge}&\: \tau \left( \sum_{n=0}^{i} {\left( 1 + 2 \mu \tau \right)}^n \right)
\end{align}
In \( (b) \) we dropped the expectation since the expression is no longer random, and \( (c) \) is the result of a geometric series. 
We return to the full expression~\eqref{eq:previous_term_5646346} and find:
\begin{align}
  &\: \mu^2 \E \Bigg \{ \mathrm{Tr}\Bigg( \boldsymbol{\Lambda}_{i^{\star}} \left( \sum_{n=0}^i \boldsymbol{D}^n \right)  \notag \\
  &\: \ \ \ \ \ \ \ \ \ \ \ \ \ \ \times \boldsymbol{V}_{i^{\star}}^{\T} \mathcal{R}_{s}\left( \bcw_{c, i^{\star}} \right) \boldsymbol{V}_{i^{\star}} \Bigg) | \w_{c, i^{\star}} \in \mathcal{H} \Bigg \} \notag \\
  \le&\: \frac{\mu}{2}  M  \sigma_u^2 - \frac{\mu}{2} \left({\left( 1 + 2\mu\tau \right)}^{i+1} - 1 \right) \sigma_{\ell}^2 \stackrel{(a)}{\le}\:-\frac{\mu}{2}M \sigma_u^2
\end{align}
where \( (a) \) holds if, and only if,
\begin{align}
  \:& \frac{\mu}{2}  M  \sigma_u^2 - \frac{\mu}{2} \left({\left( 1 + 2\mu\tau \right)}^{i+1} - 1 \right) \sigma_{\ell}^2 \le -\frac{\mu}{2}M \sigma_u^2\notag \\
  \Longleftrightarrow \:& 2 M  \frac{\sigma_u^2}{\sigma_{\ell}^2} + 1 \le {\left( 1 + 2\mu\tau \right)}^{i+1}  \notag \\
  \Longleftrightarrow \:& \log\left(2 M  \frac{\sigma_u^2}{\sigma_{\ell}^2} + 1\right) \le (i+1){\log{\left( 1 + 2\mu\tau \right)}} \notag \\
  \Longleftrightarrow \:& \frac{\log\left(2 M  \frac{\sigma_u^2}{\sigma_{\ell}^2} + 1\right)}{\log{\left( 1 + 2\mu\tau \right)}} \le {i+1} \notag \\
  \Longleftrightarrow \:& \frac{\log\left(2 M  \frac{\sigma_u^2}{\sigma_{\ell}^2} + 1\right)}{O(\mu\tau)} \le {i+1}
\end{align}
where the last line follows from \( \lim_{x \to 0} 1/x\log(1+x) = 1 \). We conclude that there exists a bounded \( i^{s} \) such that:
\begin{align}
  &\: \mu^2 \E \left\{ \mathrm{Tr}\left( \boldsymbol{\Lambda}_{i^{\star}} \left( \sum_{n=0}^{i^s} \boldsymbol{D}^n \right) \boldsymbol{V}_{i^{\star}}^{\T} \mathcal{R}_{s}\left( \bcw_{c, i^{\star}} \right) \boldsymbol{V}_{i^{\star}} \right) \right \} \notag \\
  \le&\: - \frac{\mu}{2} M \sigma_u^2
\end{align}
Applying this relation to~\eqref{eq:centered_error_recursion} and {taking expectations over \( \w_{c, i^{\star}} \in \mathcal{H} \), we obtain:}
\begin{align}
  &\: \E\left \{ {\left \| \check{\w}'{}^{i^{\star}}_{i^s+1} \right \|}_{\boldsymbol{\Lambda}_{i^{\star}}}^2 | \w_{c, i^{\star}} \in \mathcal{H} \right\} \notag \\
  \le&\: \mu^2 \sum_{n=0}^{i^s} \E\left \{ \left( \mathrm{Tr}\left( \boldsymbol{\Lambda}_{i^{\star}} \boldsymbol{D}^n \right) \cdot \E \left \{ \boldsymbol{q}_{i^{\star}+n} | \boldsymbol{\mathcal{F}}_{i^{\star}} \right \} \right) | \w_{c, i^{\star}} \in \mathcal{H} \right\} \notag \\
  &\: - \frac{\mu}{2} M \sigma_u^2
\end{align}
We now bound the perturbation term:
\begin{align}
  &\:\mu^2 \sum_{n=0}^{i^s} \E\left \{ \left( \rho\left( \boldsymbol{\Lambda}_{i^{\star}} \boldsymbol{D}^n \right) \cdot \E \left \{ \boldsymbol{q}_{i^{\star}+n} | \boldsymbol{\mathcal{F}}_{i^{\star}} \right \} \right) | \w_{c, i^{\star}} \in \mathcal{H} \right \} \notag \\
  \le&\:\mu^2 \sum_{n=0}^{i^s} \E \left \{ \left( \rho\left( \delta I {\left( I + 2 \mu \delta I \right)}^n \right) \cdot \E \left \{ \boldsymbol{q}_{i^{\star}+n} | \boldsymbol{\mathcal{F}}_{i^{\star}} \right \} \right) | \w_{c, i^{\star}} \in \mathcal{H} \right \} \notag \\
  =&\:\mu^2 \sum_{n=0}^{i^s} \left( \delta {\left( 1 + 2 \mu \delta  \right)}^n  \cdot \E \left\{ \boldsymbol{q}_{i^{\star}+n}  | \w_{c, i^{\star}} \in \mathcal{H} \right\} \right)\notag \\
  \stackrel{\eqref{eq:perturbation_definition}}{=}&\:\mu^2 \sum_{n=0}^{i^s} \delta {\left( 1 + 2 \mu \delta  \right)}^n \cdot \Bigg( \beta_R p_{\max}\Big( \E \left\{ {\left \| \widetilde{\w}^{{i}^{\star}}_{i}\right \|}^{\gamma}| \w_{c, i^{\star}} \in \mathcal{H} \right \} \notag \\
  &\: \ \ \ + \E \left \{ \left\|\bcw_{c, i^{\star}+i} - \bcw_{i^{\star}+i}\right\|^{\gamma} | \w_{c, i^{\star}} \in \mathcal{H} \right \} \Big) \notag \\
  &\: \ \ \ + \delta^2 \E \left \{ {\left \| \check{\w}'{}^{i^{\star}}_{i} \right \|}^2 | \w_{c, i^{\star}} \in \mathcal{H} \right \} \Bigg) \notag \\
  \ifarx \le&\:\mu^2 \sum_{n=0}^{i^s} \delta {\left( 1 + 2 \mu \delta  \right)}^n \cdot \left( O(\mu^{\gamma}) + \frac{O(\mu^{\gamma})}{\pi_{i^{\star}}^{\mathcal{H}}} + O(\mu^2) \right)  \notag \\ \fi
  \le&\: \delta \left( \sum_{n=0}^{i^s} {\left( 1 + 2 \mu \delta  \right)}^n\right) \left( O(\mu^{2+\gamma}) + \frac{O(\mu^{2+\gamma})}{\pi_{i^{\star}}^{\mathcal{H}}} \right) \notag \\
  \stackrel{(a)}{\le}&\: O(\mu^{1+\gamma}) + \frac{O(\mu^{1+\gamma})}{\pi_{i^{\star}}^{\mathcal{H}}} =\: o(\mu) + \frac{o(\mu)}{\pi_{i^{\star}}^{\mathcal{H}}}
\end{align}
where \( (a) \) follows from Lemma~\cite[Lemma 3]{Vlaski19nonconvexP1}. We conclude:
\begin{align}
  \E \left\{{\left \| \check{\w}'{}^{i^{\star}}_{i^s+1} \right \|}_{\boldsymbol{\Lambda}_{i^{\star}}}^2 | \w_{c, i^{\star}} \in \mathcal{H} \right\} \le  - \frac{\mu}{2} M \sigma_u^2 + o(\mu) + \frac{o(\mu)}{\pi_{i^{\star}}^{\mathcal{H}}} \label{eq:centered_deviation_bound}
\end{align}
Returning to~\eqref{eq:t_step_descent_simplified}, we find:
\begin{align}
  &\: \E \left \{ J(\w_{c, i^{\star}+i}') | \w_{c, i^{\star}} \in \mathcal{H} \right\} \notag \\
  \le&\: \E \left \{J(\w_{c, i^{\star}}) | \w_{c, i^{\star}} \in \mathcal{H} \right\} + \frac{1}{2} \E \left \{ {\left \| \check{\w}'{}^{i^{\star}}_{i} \right \|}_{\boldsymbol{\Lambda}_{i^{\star}}}^2 | \w_{c, i^{\star}} \in \mathcal{H} \right \} \notag \\
  &\: + \frac{\rho}{6} \E \left \{ {\left \| \widetilde{\w}_{i}'{}^{i^{\star}} \right \|}^3 | \w_{c, i^{\star}} \in \mathcal{H} \right \} \notag \\
  \le&\: \E \left \{J(\w_{c, i^{\star}}) | \w_{c, i^{\star}} \in \mathcal{H} \right\} - \frac{\mu}{2} M \sigma_u^2 + o(\mu) + \frac{o(\mu)}{\pi_{i^{\star}}^{\mathcal{H}}}
\end{align}

\section{Proof of Theorem~\ref{TH:FINAL_THEOREM}}\label{AP:FINAL_THEOREM}
The proof follows by constructing a particular telescoping sum and subsequently applying~\cite[Theorem 2]{Vlaski19nonconvexP1} and~\ref{TH:DESCENT_THROUGH_SADDLE_POINTS}. To begin with, we define the stochastic process:
\begin{equation}
  \mathbf{t}(k+1) = \begin{cases} \mathbf{t}(k) + 1, \ &\mathrm{if} \ \w_{c, \mathbf{t}(k)} \in \mathcal{G}, \\\mathbf{t}(k) + 1, \ &\mathrm{if} \ \w_{c, \mathbf{t}(k)} \in \mathcal{M}, \\ \mathbf{t}(k) + i_s, \ &\mathrm{if} \ \w_{c, \mathbf{t}(k)} \in \mathcal{H}. \end{cases}
\end{equation}
where \( \mathbf{t}(0) = 0 \). We then have:
\begin{align}
  &\: \E \left \{ {J(\w_{c, \mathbf{t}(k)}) - J(\w_{c, \mathbf{t}(k+1)})} | \w_{c, \mathbf{t}(k)} \in \mathcal{G} \right \} \notag \\
  =&\: \E \left \{ {J(\w_{c, \mathbf{t}(k)}) - J(\w_{c, \mathbf{t}(k)+1})} | \w_{c, \mathbf{t}(k)} \in \mathcal{G} \right \} \notag \\
  \stackrel{}{\ge}&\: \mu^2 \frac{c_2}{\pi} - O(\mu^3) - \frac{O(\mu^3)}{\pi_i^{\mathcal{G}}}
\end{align}
and
\begin{align}
  &\: \E \left \{ {J(\w_{c, \mathbf{t}(k)}) - J(\w_{c, \mathbf{t}(k+1)})} | \w_{c, \mathbf{t}(k)} \in \mathcal{H} \right \} \notag \\
  =&\: \E \left \{ {J(\w_{c, \mathbf{t}(k)}) - J(\w_{c, \mathbf{t}(k)+1})} | \w_{c, \mathbf{t}(k)} \in \mathcal{H} \right \} \notag \\
  {\ge}&\: \frac{\mu}{2} M \sigma_u^2 - o(\mu) - \frac{o(\mu)}{\pi_i^{\mathcal{H}}}
\end{align}
Finally, we have:
\begin{align}
  &\: \E \left \{ {J(\w_{c, \mathbf{t}(k)}) - J(\w_{c, \mathbf{t}(k+1)})} | \w_{c, \mathbf{t}(k)} \in \mathcal{M} \right \} \notag \\
  =&\: \E \left \{ {J(\w_{c, \mathbf{t}(k)}) - J(\w_{c, \mathbf{t}(k)+1})} | \w_{c, \mathbf{t}(k)} \in \mathcal{M} \right \} \notag \\
  \stackrel{}{\ge}&\: - \mu^2 c_2 - O(\mu^3) - \frac{O(\mu^3)}{\pi_i^{\mathcal{M}}}
\end{align}
where \( (a) \) follows since \( \mathbf{t}(k+1) - \mathbf{t}(k) = 1 \) when \( \w_{c, \mathbf{t}(k)} \in \mathcal{M} \). We can combine these relations to obtain:
\begin{align}
  &\: \E \left \{ {J(\w_{c, \mathbf{t}(k)}) - \E J(\w_{c, \mathbf{t}(k+1)})} \right \} \notag \\
  =&\: \E \left \{ {J(\w_{c, \mathbf{t}(k)}) - \E J(\w_{c, \mathbf{t}(k+1)})} | \w_{c, \mathbf{t}(k)} \in \mathcal{G} \right \} \cdot \pi_{\mathbf{t}(k)}^{\mathcal{G}} \notag \\
  &\:+ \E \left \{ {J(\w_{c, \mathbf{t}(k)}) - \E J(\w_{c, \mathbf{t}(k+1)})} | \w_{c, \mathbf{t}(k)} \in \mathcal{H} \right \} \cdot \pi_{\mathbf{t}(k)}^{\mathcal{H}} \notag \\
  &\:+ \E \left \{ {J(\w_{c, \mathbf{t}(k)}) - \E J(\w_{c, \mathbf{t}(k+1)})} | \w_{c, \mathbf{t}(k)} \in \mathcal{M} \right \} \cdot \pi_{\mathbf{t}(k)}^{\mathcal{M}} \notag \\
  =&\: \left( \mu^2 \frac{c_2}{\pi} - O(\mu^3) - \frac{O(\mu^3)}{\pi_i^{\mathcal{G}}} \right) \cdot \pi_{\mathbf{t}(k)}^{\mathcal{G}} \notag \\
  &\:+ \left( \mu \frac{c_2}{\pi i_s} - o(\mu) - \frac{o(\mu)}{\pi_i^{\mathcal{H}}} \right) \cdot \pi_{\mathbf{t}(k)}^{\mathcal{H}} \notag \\
  &\:+ \left( - \mu^2 c_2 - O(\mu^3) - \frac{O(\mu^3)}{\pi_i^{\mathcal{M}}} \right) \cdot \pi_{\mathbf{t}(k)}^{\mathcal{M}} \notag \\
  =&\: \mu^2 \frac{c_2}{\pi} \cdot \pi_{\mathbf{t}(k)}^{\mathcal{G}} + \left(\frac{\mu}{2} M \sigma_u^2 - o(\mu) \right) \cdot \pi_{\mathbf{t}(k)}^{\mathcal{H}} \notag \\
  &\: - \mu^2 c_2 \cdot \pi_{\mathbf{t}(k)}^{\mathcal{M}} - o(\mu^2)
\end{align}
Suppose \( \pi_{\mathbf{t}(k)}^{\mathcal{M}} \le 1 - \pi \) for all \( i \). Then \( \pi_{\mathbf{t}(k)}^{\mathcal{G}} + \pi_{\mathbf{t}(k)}^{\mathcal{H}} \ge \pi \) for all \( i \), and
\begin{align}
  &\: \E \left \{ {J(\w_{c, \mathbf{t}(k)}) - \E J(\w_{c, \mathbf{t}(k+1)})} \right \} \notag \\
  \ge&\: \mu^2 \frac{c_2}{\pi} \cdot \left( \pi - \pi_{\mathbf{t}(k)}^{\mathcal{H}} \right)+ \left( \frac{\mu}{2} M \sigma_u^2 - o(\mu) \right) \cdot \pi_{\mathbf{t}(k)}^{\mathcal{H}} \notag \\
  &\: - \mu^2 c_2 \cdot \left( 1 - \pi \right) - o(\mu^2) \notag \\
  =&\: \mu^2 {c_2}\pi  +\left( \frac{\mu}{2} M \sigma_u^2 - \mu^2 \frac{c_2}{\pi} - o(\mu)\right)  \pi_{\mathbf{t}(k)}^{\mathcal{H}} - o(\mu^2) \notag \\
  \stackrel{(a)}{\ge}&\: \mu^2 {c_2}\pi - o(\mu^2)
\end{align}
where \( (a) \) holds whenever \(  \frac{\mu}{2} M \sigma_u^2 - \mu^2 \frac{c_2}{\pi} - o(\mu) \ge 0 \), which holds whenever \( \mu \) is sufficiently small. We hence have by telescoping:
\begin{align}
  &\: J(w_{c, 0}) - J^o \notag \\
  \ge&\: \E J(w_{c, \mathbf{t}(0)}) - \E J(\w_{c, \mathbf{t}(k)}) \notag \\
  =&\: \E J(w_{c, \mathbf{t}(0)}) - \E J(\w_{c, \mathbf{t}(1)}) \notag \\
  &\: + \E J(\w_{c, \mathbf{t}(1)}) - \E J(\w_{c, \mathbf{t}(2)}) \notag \\
  &\: + \cdots \notag \\
  &\: + \E J(\w_{c, \mathbf{t}(k-1)}) - \E J(\w_{c, \mathbf{t}(k)}) \notag \\
  \ge&\: \mu^2 c_2 \pi k
\end{align}
Rearranging yields:
\begin{equation}
  k \le \frac{J(w_{c, 0}) - J^o}{\mu^2 c_2 \pi}
\end{equation}
We conclude by definition of the stochastic process \( \boldsymbol{t}_k \):
\begin{equation}
  i = \boldsymbol{t}(k) \le k \cdot i^s \le \frac{\left( J(w_{c, 0}) - J^o \right)}{\mu^2 c_2 \pi} i^s
\end{equation}

%
\bibliographystyle{IEEEbib}
\bibliography{nonconvex}

\end{document}